\documentclass[twocolumn]{aastex631}

\usepackage{savesym}
\savesymbol{tablenum}
\usepackage{siunitx}
\restoresymbol{SIX}{tablenum}
\usepackage{multirow}

\usepackage{wasysym}

\DeclareSIUnit\au{au}
\DeclareSIUnit\msun{M_\odot}
\DeclareSIUnit\mearth{M_\oplus}
\DeclareSIUnit\year{yr}
\DeclareSIUnit\erg{erg}
\DeclareSIUnit\eV{eV}

\received{2023-03-20}
\revised{2023-12-24}
\accepted{2023-12-31}
\submitjournal{the Planetary Science Journal}

\shorttitle{A New Database of Giant Impacts}
\shortauthors{Emsenhuber et al.}

\begin{document}

\title{A New Database of Giant Impacts over a Wide Range of Masses and with Material Strength:\\ A First Analysis of Outcomes}

\correspondingauthor{Alexandre Emsenhuber}
\email{alexandre.emsenhuber@unibe.ch}

\author[0000-0002-8811-1914]{Alexandre Emsenhuber}
\affiliation{Universitäts-Sternwarte, Ludwig-Maximilians-Universität München, Scheinerstraße 1, 81679 München, Germany}
\affiliation{Lunar and Planetary Laboratory, University of Arizona, 1629 E. University Blvd., Tucson, AZ 85721, USA}
\affiliation{Space Research and Planetary Science, University of Bern, Gesellschaftsstrasse 6, 3012 Bern, Switzerland}

\author[0000-0003-1002-2038]{Erik Asphaug}
\affiliation{Lunar and Planetary Laboratory, University of Arizona, 1629 E. University Blvd., Tucson, AZ 85721, USA}

\author[0000-0001-6294-4523]{Saverio Cambioni}
\affiliation{Department of Earth, Atmospheric \& Planetary Sciences, Massachusetts Institute of Technology, 77 Massachusetts Avenue, Cambridge, MA 02139, USA}

\author[0000-0002-9767-4153]{Travis S. J. Gabriel}
\affiliation{U.S. Geological Survey, Astrogeology Science Center, 2255 N. Gemini Drive, Flagstaff, AZ 86001, USA}

\author[0000-0001-5475-9379]{Stephen R. Schwartz}
\affiliation{Lunar and Planetary Laboratory, University of Arizona, 1629 E. University Blvd., Tucson, AZ 85721, USA}
\affiliation{Planetary Science Institute, 1700 E.\ Fort Lowell, Suite 106, Tucson, AZ 85719-2395, USA}

\author[0000-0003-2018-3273]{Robert E. Melikyan}
\affiliation{Lunar and Planetary Laboratory, University of Arizona, 1629 E. University Blvd., Tucson, AZ 85721, USA}

\author[0000-0002-6696-2961]{C. Adeene Denton}
\affiliation{Lunar and Planetary Laboratory, University of Arizona, 1629 E. University Blvd., Tucson, AZ 85721, USA}

\begin{abstract}
In the late stage of terrestrial planet formation, planets are predicted to undergo pairwise collisions known as giant impacts. Here we present a high-resolution database of giant impacts for differentiated colliding bodies of iron-silicate composition, with target masses ranging from \SI{e-4}{\mearth} up to super-Earths (\SI{5}{\mearth}). We vary impactor-to-target mass ratio, core-mantle (iron-silicate) fraction, impact velocity, and impact angle. Strength in the form of friction is included in all simulations. We find that due to strength, collisions with bodies smaller than about \SI{2e-3}{\mearth} can result in irregular shapes, compound core structures, and captured binaries. We observe that the characteristic escaping velocity of smaller remnants (debris) is approximately half of the impact velocity, significantly faster than currently assumed in \textit{N}-body simulations of planet formation. Incorporating these results in \textit{N}-body planet formation studies would provide more realistic debris-debris and debris-planet interactions.
\end{abstract}

\section{Introduction}
\label{sec:intro}

In the last stage of classical terrestrial planet formation, collisions between similar-sized planetary embryos are thought to be the dominant mode of growth \citep[e.g.,][]{1985ScienceWetherill,2002ApJKokubo} where Moon- to Mars-sized bodies accumulate dynamically to form the final planets. Other stages of planet, planetesimal, and satellite formation may also involve giant impacts, or more generally, similar-sized collisions \citep{2010ChEGAsphaug}. A correct understanding of how planets accumulate and exchange matter in these numerous giant impacts thus underlies our most basic knowledge of planet formation.

Giant impacts are complicated phenomena. Colliding bodies can be centrally condensed, leading to large mass fractions outside the direct collision zone \citep[e.g.,][]{2012ApJGenda,2016IcarusMovshovitz}. Off-axis impacts involve high angular momentum and limited accretion efficiency \citep{2004ApJAgnor}. They result in a complicated post-collision phase \citep[e.g.,][]{1997IcarusCameron}. The fraction of the impactor mass $m_\mathrm{imp}$ that gets accreted onto the target of original mass $m_\mathrm{tar}$ and final mass $m_\mathrm{L}$ is given by the accretion efficiency
\begin{equation}
\label{eq:acc_L}
    \xi_\mathrm{L} = \frac{m_\mathrm{L} - m_\mathrm{tar}}{m_\mathrm{imp}}.
\end{equation}
A perfect merger has the largest possible accretion efficiency, $\xi_\mathrm{L}=1$. Net accretion requires $\xi_\mathrm{L}>0$, and $\xi_\mathrm{L}<0$ describes mass loss (erosion or disruption).

The slowest possible collisions occur at around the mutual escape velocity at contact, where for colliding spheres
\begin{equation}
    v_\mathrm{esc} = \sqrt{\frac{2G(m_\mathrm{tar}+m_\mathrm{imp})}{r_\mathrm{tar}+r_\mathrm{imp}}}.
\end{equation}
Here $G$ is the gravitational constant and $r_\mathrm{tar}$ and $r_\mathrm{imp}$ are the target and impactor radii, respectively. The largest bodies of a growing planetary system, under conditions of gravitational self-stirring, collide at relative velocities near their mutual escape velocity \citep[e.g.,][]{1985ScienceWetherill}. \textit{N}-body simulations find that most solar system giant impacts are faster than $v_\mathrm{esc}$ ($\sim$1--4~$v_{\rm{esc}}$, e.g., \citealp{1999IcarusAgnor,2016ApJQuintana}) and the impact angle is often off-axis. Because of this, the impactor and target can undergo `hit-and-run', where the bodies remain relatively unscathed after a glancing blow and accretion efficiency is close to zero \citep{2004ApJAgnor}. This is expected to regulate the pace/velocity of planet formation. In these typical scenarios, the impactor plows through the target mantle and emerges as a deflected, decelerated, and gravitationally-intact body called the ``runner''. The bodies may then recollide after orbiting the central star.

At more head-on and/or at lower velocities ``graze-and-merge'' collisions are possible \citep[e.g.,][]{2010ApJLeinhardt}. These are high angular momentum accretions, as represented by the canonical Moon-forming giant impact \citep{2001NatureCanup}. 
When averaged over impact angle, graze-and-merge may be the dominant form of giant impact accretion \citep{2012ApJStewart}.

In some graze-and-merge-like scenarios, the gravitationally-bound runner can overshoot the target, but the bodies may not entirely escape one another. Tidal friction and transfer of angular momentum around an irregular central mass can then cause material to be captured as a moon, at least temporarily; such ``graze-and-capture'' collisions include hypotheses for Moon formation \citep{1987IcarusBenz} and Pluto-Charon formation \citep{2005ScienceCanup}. This represents an intermediate case between graze-and-merge and hit-and-run.

\subsection{Dynamical significance}

The simulation of the gravitational interactions of a planetary system (\textit{N}-body simulations) depends on how collisions are treated. If colliding bodies are assumed to be perfectly merging, mass is conserved and the new orbit is often placed at the center of mass of the colliding bodies \citep[e.g.,][]{1998AJDuncan}. Perfect merging may be a sufficient assumption for understanding the largest-scale architectures in the solar system \citep[e.g.,][]{2010ApJKokubo,2016ApJQuintana,2019IcarusWalshLevison}. However, the approximation alters the dynamical evolution and the formation sequence.

Including realistic collision outcomes increases the formation timescales \citep{2013IcarusChambers,2016ApJQuintana}, because inefficient accretions---especially hit-and-runs---are common. If the runner returns to the target, or to another nearby accreting body \citep{2021PSJEmsenhuber}, it is a ``collision chain'', a hit-and-run followed by another similar-sized collision (merger, hit-and-run, disruption, etc.). During planet formation at \SI{\sim1}{\au} around a solar-mass star, the recollision time for a collision chain can be \num{e3}--\SI{e6}{\year} \citep{2019ApJEmsenhuberA}. Such multi-collisional pathways could lead to mantle-stripped cores, the ``stranded runner'' hypothesis for the origin of Mercury \citep{2014NatGeoAsphaug,2018ApJChau}, and for metallic asteroids and meteorites \citep{2007NatureYang}.

Also, a realistic treatment of collisions affects the resulting composition \citep{2015IcarusDwyer,2015ApJCarter}, and the mixing of material between planetary accretion zones \citep{2020AABurger}, and leaves behind a diversity of smaller bodies (e.g., Mars-sized; \citealp{2020ApJEmsenhuberA}). This is because a growing fraction of the remainder end up surviving hit-and-run collisions as growth of the largest bodies proceeds \citep{2014NatGeoAsphaug,2017BookAsphaug}. This could explain the observed increasing diversity of terrestrial planets with decreasing mass \citep{2023AREPSGabrielCambioni}.

Implementing realistic treatments of similar-sized collisions can be achieved in \textit{N}-body codes in several ways. For example, scaling laws have been developed \citep[e.g.,][]{2012ApJLeinhardt,2022MNRASReinhardt} from 3D giant impact simulations that characterize the outcomes of giant impacts. 
One application of our new database would be to validate or update the parameters to such scaling laws or develop new ones, as our parameter space includes friction in all simulations, providing more accurate results for impacts at smaller scales. 
Another approach is to apply machine learning to the outcomes of giant impact simulations \citep{2020CACTimpe}. Surrogate models \citep[e.g.,][]{2019ApJCambioni,2021PSJCambioni} can be generated, which relate pre-impact conditions to post-impact conditions such as the largest remnant mass $m_\mathrm{L}$, the second remnant mass $m_\mathrm{S}$ (runner), dynamical properties, and compositional information. The present database, including an initial assessment of debris, is designed specifically for building machine learning models and spans a larger range of collisions than an existing giant impact database used for machine learning models \citep{2019ApJCambioni,2020ApJEmsenhuberA}. 
One important limitation, however, is that impacting bodies are not rotating before impact in our database. Introducing rotation would expand the parameter space considerably, but is a factor to consider given its effect on post-impact outcomes \citep{2020CACTimpe}, and the profound effects it may have on the internal structure of the resulting bodies \citep{2017JGRELockStewart}.

\subsection{Geophysical significance}

Certain aspects of a giant impacts, such as the mass of remnants, can be modeled by analytical relationships (such as scaling laws). However, obtaining a unified model that can predict the outcomes of impacts across various regimes, such as in small asteroid-scale collisions and in shock-inducing Moon-formation events, is challenging. This is due to numerous physical complexities inherent to collisions at different scales and in different bodies \citep{2020ApJGabriel}. Transitions in the equations of state (EOSs) \citep{2019AIPStewart}, vapor production \citep{2007SSRvBenz,2020JGRECarter,2020JGREDavies}, and the presence of shocks in large-scale collisions produce ample vapor and may alter the nature of erosion \citep{2020ApJGabriel,2021ApJGabrielAllenSutter}. At small scales, friction and strength make erosion less likely for a given scaled velocity \citep[e.g.,][]{2015PSSJutzi} and other forms of dissipation become substantial \citep{2018GRLMeloshIvanov}. Even in Mars-scale collisions where rheology has not classically been noted to alter the mass of the largest remnant \citep[e.g.,][]{1999IcarusBenzAsphaug}, other aspects of the collision such as heating and debris generation are influenced by the presence of strength \citep{2018IcarusEmsenhuber}. To make meaningful progress towards a unifying model across these complex regimes, a set of simulations that spans over a large range of pre-impact conditions is required.

Still, giant impact outcomes do have systematic trends across vast ranges. For instance, \citet{2019PSSJutzi} identified three basic regimes of giant impacts, or similar-sized collisions, for the velocity ranges that can potentially result in accretion. 
For bodies smaller than \SI{\sim100}{\kilo\meter} there is the porosity regime, where the outcome is mainly affected by pore crushing \citep[e.g.,][]{1999IcarusHousenHolsapple,2007IcarusBelton,2015AstIVJutzi}. Intermediate-mass bodies (up to \SI{\sim1000}{\kilo\meter}), are in the strength regime, where friction is important (for the link between strength and friction, see Section \ref{sec:strength}). The largest bodies are in the gravity regime, where strength and porosity can be ignored.
For large enough bodies where mutual escape velocities (and thus impact velocities) exceed the sound speed, shocks dominate result in a fourth regime \citep{2020ApJGabriel}.
These regimes are idealizations, and the transitions are not abrupt. Indeed, the transition from strength to gravity dominance may span the entire range of ``oligarchic growth'', the canonical late stage of Moon- to Mars-sized bodies \citep{2002ApJKokubo}. For this reason, we include a strength model in all of our simulations, even in super-Earth collisions where the effect on the outcome is expected to be negligible. This allows the data to properly capture the smooth transition from strength to gravity dominance.

The end state of the last giant impact represents the beginning of a planet's long-term geophysical evolution \citep[e.g.,][]{2007SSRvZahnle}.
Accreted planets can end up with mantles and cores unstable to convection, especially for larger scale collisions in which the gravitational potential released by the merging material exceeds the energy of melting \citep{2020EPSLLock}. Smooth Particle Hydrodynamics (SPH), the technique most widely used to model giant impacts, is unable to properly represent long-term convective instability and other effects that happen on a much longer timescale than the collision.
To study post-impact geodynamics requires a hybrid numerical approach, as has been applied by \citet{2018IcarusGolabek} to model consequences like geothermal overturn, geochemistry and solidification after giant impacts.
For solid final bodies, finite element analysis could be used to predict post-collisional relaxation from these SPH results.

\subsection{Geochemical significance}

Giant impacts are transformative events, and simulations of giant impacts have become the basis for making quantitative geochemical predictions about late stage planet formation \citep[e.g.,][]{2018ApJHaghighipour}, especially the origin of the Moon \citep{2021BookCanup}. 
But colliding planetary bodies in simulations are represented by idealized compositions, and these and other simplifying assumptions need to be recognized. 
We represent terrestrial planets and their progenitors as differentiated nonporous spheres of forsterite and iron composition, with varying core mass fractions. 
This simple interior structure allows us to focus on attaining accuracy and reliability of the model predictions, while making a sufficient sweep of the parameter space to obtain results that may be applied generally to models of terrestrial planet formation. Furthermore, there remains limited availability of shock EOS models for a wider range of mantle materials and the forsterite EOS \citep{2019AIPStewart} is most up-to-date and widely used for our purposes.

A two-component rock/iron interior is a justifiable approximation for terrestrial bodies in the modeled size range. For instance, 525-km diameter (4) Vesta has a rocky mantle and an iron core \citep[e.g.,][]{2012ScienceRussell}; so do Mercury, Venus, the Moon, and Earth, to good approximation. So we maintain this assumption from the smallest colliding bodies in our database, about 1000-km diameter, up to \SI{\sim5}{\mearth}, beyond which point gas-free accretion is unlikely \citep[e.g.,][]{2015ApJRogers}. 
 
Numerical studies have revealed the complex thermodynamic evolution that occurs in giant impacts \citep{2020JGRECarter} and how outcomes can depend on the choice of initial thermal conditions, even for relatively simple two-component planets. For example, a massive magma ocean on the target can enhance the post-impact disk mass and its Earth-isotopic fraction in simulations of Moon formation \citep{2019NatGeoHosono}. That said, the material from which the Moon forms represents only \SI{\sim1}{\percent} of the material of the colliding bodies in the Canonical model for Moon formation \citep{2005ScienceCanup}. For our database we thus implement only one thermal state across all colliding bodies: both the core and the mantle start slightly below the solidus, on the basis that convective cooling may be faster than the time between giant impacts. 

\subsection{Scope of work}

We present a significant new database of terrestrial planet-forming giant impacts. To take into account the limitations and bottlenecks of previous works, our new database has the following characteristics:
\begin{itemize}
\item It is applicable to a wide range of impactor-target properties and impact parameters, from the sizes of the largest asteroids and rocky satellites, to terrestrial planets of several Earth-masses (\si{\mearth}), which are not expected to be large enough to accrete significant atmospheres \citep[e.g.,][]{2015ApJRogers}.
\item It includes transitions between collision regimes, especially between graze-and-merge and hit-and-run, i.e. accretion and non-accretion.
\item It provides the outcomes of collisions for bodies with various core-mass-fractions, thus enabling self-consistent treatments of core size evolution in collision chains.
\end{itemize}

Within this scope we construct a database of \num{1250} simulations (split in two sets, one comprising \num{1000} simulations and one \num{250}) suitable for the development of machine learning models.
To compute the final reduced properties of each simulation, we proceed in multiple steps.
First, we define the setup of the study (Section~\ref{sec:methods-params}) and determine the initial conditions of hydrodynamical simulations (Section~\ref{sec:methods-range}).
Then, we perform simulations using SPH (Section~\ref{sec:methods-sims}) and analyze their results (Section~\ref{sec:methods-sims}). 
We also present several results from our calculation: general results (Section~\ref{sec:res-gen}), specific items relevant to low-velocity grazing collisions: satellite capture (Section~\ref{sec:res-sat}), body shapes (Section~\ref{sec:res-shape}), and debris (Section~\ref{sec:res-debris}).

\section{Methods}
\label{sec:methods}

\subsection{Studied parameters}
\label{sec:methods-params}

The first step to generating our database of giant impact simulations is to select the parameters that will be explored. Target mass $m_\mathrm{tar}$ and the impactor-to-target mass ratio $\gamma=m_\mathrm{imp}/m_\mathrm{tar}$ are the primary parameters. For the composition, we use a single parameter, the core mass fraction $Z_\mathrm{tar}$ and $Z_\mathrm{imp}$, which describes the fraction of iron in our two-layered bodies, the rest being forsterite. Core mass fraction is a free parameter in our database as it will allow us to understand collision outcomes between planets that have been previously eroded, e.g., to know the outcomes of collisions between core-rich bodies. The impact velocity is given in terms of the mutual escape velocity $v_\mathrm{coll}/v_\mathrm{esc}$, which thus increases with total mass. The impact angle $\theta_\mathrm{coll}$ is defined as the angle between the line connecting the centers of mass and the relative velocity vector at impact. We use the convention that $\theta_\mathrm{coll}=\ang{0}$ is a head-on collision and $\theta_\mathrm{coll}=\ang{90}$ is a grazing collision.

For the present work we have decided not to include the effect of pre-impact rotation. \citet{2020CACTimpe} found that pre-impact rotation had less influence on the giant impact outcomes (apart from final spin, which is strongly correlated) than the colliding mass, mass ratio, core mass fraction, impact angle, and impact velocity. An important advantage of leaving out pre-impact rotation is the smaller dimensionality of the parameter space. Accounting for all possibilities of pre-impact rotation when the target and impactor are similar in size would require six additional parameters, three for each body, to describe the spin angular momentum vectors.

\subsection{Parameter range and initial conditions}
\label{sec:methods-range}

We aim to study the effect of the target's mass, so we explore a large range from \SI{e-4}{\mearth} to \SI{5}{\mearth}, which is sampled uniformly in logarithm space. The lower boundary of target mass corresponds to about the mass of (1)~Ceres. This range was selected because it incorporates the size range where the effect of friction (included in every simulation in the generated database) is important \citep[e.g.,][]{2018IcarusEmsenhuber}, yet where porosity is likely unimportant for these massive, differentiated bodies. 
The upper boundary was selected so that the collisions can be applied to the formation of extrasolar planetary systems that contain super-Earths. We do not include simulations of planets with gas envelopes, so we limit our study to masses below \SI{5}{\mearth} because more massive planets tend to have significant envelopes \citep[e.g.,][]{2015ApJRogers}.

As for the mass ratio $\gamma$, our sampling is uniform so that unequal mass ratios are studied as much as more similar mass ratios, as suggested by \citet{2019ApJValencia}. We selected the range of $0.05 < \gamma < 1$, covering the $\sim1:3$ diameter range defining similar-sized collisions \citep{2010ChEGAsphaug} and as constrained by our capability to computationally resolve the smaller body. The least massive impactor in our database is \SI{5e-6}{\mearth}, somewhat more massive than Main Belt asteroid (16)~Psyche.

Core mass fractions $Z$ are sampled from a piecewise uniform distribution over the range of \SI{10}{\percent} to \SI{90}{\percent}, where the range is selected so that the core and mantle are numerically resolved at least several particles across for the chosen resolution. Further, to account for the population of bodies around the average value of Fe/Mg for stars with planets from the Hypatia Catalogue \citep[see also Figure 2 in \cite{2020MNRASScora}]{2014AJHinkel}, and the average chondritic values of the terrestrial planets, we center the distribution so that half the bodies have a core mass fraction below \SI{30}{\percent} and the rest above. For this median value, which is around the value for Earth and Venus, the core radius is about half the body radius.

\begin{table}
    \centering
    \caption{Cumulative distribution of impact velocities in the grid of hydrodynamical simulations.}
    \label{tab:vel-dist}
    \begin{tabular}{ccc}
        $v_\infty/v_\mathrm{esc}$ & $v_\mathrm{coll}/v_\mathrm{esc}$ & Cumulative \\
        & & fraction \\
        \hline
        0 & 1 & 0 \\
        1 & \num{\sim1.4} & 0.4 \\
        2 & \num{\sim2.2} & 2/3 \\
        3.5 & \num{\sim3.6} & 0.8 \\
        6 & \num{\sim6.1} & 0.9 \\
        $\sqrt{99}$ & 10 & 1 \\
    \end{tabular}

\end{table}

\begin{figure}
    \centering
    \includegraphics{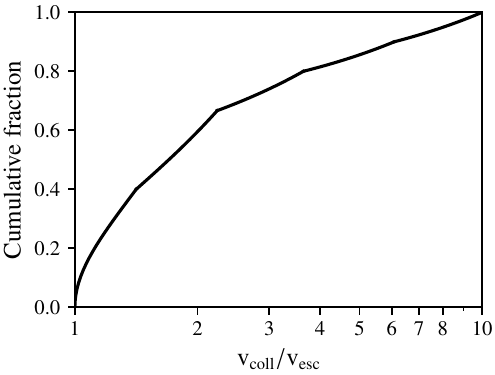}
    \caption{Cumulative distribution of impact velocities in our giant impact database. The parameter distribution was chosen to sample regions where post-impact outcomes transition from one type of impact outcome (e.g., accretion) to another (e.g., hit-and-run).}
    \label{fig:vel-dist}
\end{figure}

According to \textit{N}-body studies, the majority of giant impacts during terrestrial planet formation occur in the range $1-2 v_\mathrm{esc}$ \citep[e.g.,][]{2013IcarusChambers,2020ApJEmsenhuberA}, although the median value depends on the damping effects of planetesimals \citep{2006IcarusOBrien}. According to \citet{2010ChEGAsphaug} the diversity of planet formation by giant impacts is due to the fact that the transitions in outcomes happen around this accretionary range of velocities. 
\citet{2012ApJGenda} noted significant variability at the low velocity merging and graze-and-merge transition, for minor changes in simulation parameters. Similarly \citet{2019ApJCambioni} found the highest rate of misclassification and prediction error to occur at the transition between hit-and-run and merging, in their machine learning analysis of a prior database of giant impact outcomes.
Therefore we place emphasis on mapping out this transition by constructing a distribution of impact velocity that favors low-velocity collisions. We note that minor deviations in the transition as a function of different thermal conditions are not covered in this work. We define the cumulative distribution from the relative velocity at infinity, as scaled by the mutual escape velocity $v_\infty/v_\mathrm{esc}$. This is related to the impact velocity by
\begin{equation}
    \left(\frac{v_\mathrm{coll}}{v_\mathrm{esc}}\right)^2=1+\left(\frac{v_\infty}{v_\mathrm{esc}}\right)^2
\end{equation}
according to energy conservation. The result is a piecewise uniform distribution as provided in Table~\ref{tab:vel-dist}. The resulting cumulative distribution, given for $v_\mathrm{coll}/v_\mathrm{esc}$, is shown in Figure~\ref{fig:vel-dist}.

As for the impact angle $\theta_\mathrm{coll}$, we follow the distribution expected for collisions between point impactors onto gravitational targets, $\mathrm{d}P\propto\sin{(2\theta_\mathrm{coll})}\mathrm{d}\theta_\mathrm{coll}$ \citep{1962BookShoemaker}. This ensures that the transitions between hit-and-run and graze-and-merge and accretion, which occur around the nominal range of impact angles at nominal velocities \citep[e.g.,][]{2012ApJLeinhardt}, is well sampled.

To generate the specific combinations of the parameters which will be used to perform the hydrodynamical simulations, we use the Latin hypercube sampling (LHS) method \citep[e.g.,][]{1979TechMcKay,2020CACTimpe}. LHS divides each parameter into $n$ intervals of equal probability based on the distribution, with $n$ being the number of samples. Then, one sample is selected randomly from each interval. This ensures that the entire range of possible values for each parameter is sampled. In addition to that, LHS adopts criteria that ensure that the entire parameter space is well sampled. For instance, the algorithm avoids correlations between parameter values in the selected samples, so that the effect of each parameter can be disentangled. In this work, we used the \texttt{pyDOE} Python package with the \texttt{minmax} setting. This packages returns all the samples in the [0,1] range with uniform probability. We convert these to the actual collision parameters using the probability distributions discussed above.

For this work, we generated two lists of initial conditions using LHS, one with $n=\num{1000}$ entries and one with $n=\num{250}$. The intention was to have separate sets for future machine learning applications. For most of the analysis, we combine the two in a set of \num{1250} simulations.

\subsection{Hydrodynamical simulations}
\label{sec:methods-sims}

To perform the hydrodynamical simulations, we use the SPH technique \citep[e.g.,][]{1992ARA&AMonaghan,2009NARRosswog}. SPH uses a Lagrangian description with continuum material divided into particles. Quantities are retrieved by performing a kernel interpolation and their spatial derivatives by interpolation of the underlying quantity with the kernel's gradient. Time evolution is given by the Euler's equations, except that density is computed by performing kernel interpolation at each step and corrected for free-surface effects using the method of \citet{2017MNRASReinhardtStadel}, which works by increasing the density of a particle if there is a spatial imbalance of neighbors around it. An artificial viscosity term inspired by Riemann solvers \citep{1997JCPMonaghan} with a time-dependent factor \citep{1997JCPMorrisMonaghan} is included to resolve shocks, the exact form of which is described in \citet{2021PSJEmsenhuber}. The interpolation is performed using a cubic spline kernel \citep{1985AAMonaghanLattanzio} with each particle having about 50 neighbors. We note that SPH tends to spuriously damp subsonic turbulence \citep[e.g.,][]{2010MNRASCullenDehnen,2012MNRASBauerSpringel,2019ApJDengA}; however this should affect more the internal mixing within the final bodies rather than their global iron-silicate fraction. Finally, material strength is modeled according the formulation discussed in the next section.

To close the Euler equations, we need to provide an EOS that provides the pressure as function of the density and the specific internal energy $p(\rho,u)$. The choice of EOS is limited by the requirement that they be thermodynamically reliable in all the energy regimes applicable to a giant impact \citep{2007M&PSMelosh}. In this work, we use \texttt{ANEOS} for the iron core \citep{ANEOS} and a modified version of \texttt{ANEOS} for Mg\textsubscript{2}SiO\textsubscript{4} (forsterite) \citep{2019AIPStewart} for the mantle. To improve performance, we precompute tables directly from ANEOS for the anticipated range of values as in previous works \citep[e.g.][]{1989IcarusBenz,2012IcarusReufer}. The code nevertheless retains the capability to call ANEOS for the few cases where the values lie outside of the tabulated range.

The computer code used in this study is \texttt{SPHLATCH} \citep[e.g.,][]{2012IcarusReufer,2013IcarusAsphaug,2014NatGeoAsphaug} and includes additional updates and corrections presented in \citet{2023IcarusBallantyne}. An earlier version of the same code was used to generate a previous collision database \citep{2011PhDReufer} on which the first machine-learning derived surrogate models were based \citep{2019ApJCambioni,2020ApJEmsenhuberA} and an updated semi-empirical scaling law was developed \citep{2020ApJGabriel}. Those simulations did not include friction and the reported database spans \num{e-2} -- \SI{1}{\mearth} planets \cite[for more information see][]{2020ApJGabriel}.

\subsection{Constitutive strength}
\label{sec:strength}

The constitutive model we adopt is similar to \citet{2018IcarusEmsenhuber}: elastic perfectly plastic material \citep{1994IcarusBenzAsphaug,1995CoPhCBenzAsphaug}, with the deformation tracked by the deviatoric stress tensor, that is reduced at the Hugoniot elastic limit with $\sqrt{J_2}/Y$, where $J_2$ is the second invariant of the deviatoric stress tensor (not to be confused with the global gravity coefficient) and $Y$ the pressure-dependent yield strength \citep{2004MaPSCollins,2015PSSJutzi}. We further include a correction tensor to compute the local velocity gradient to achieve angular momentum conservation \citep{1999CMABonet,2006HabilSpeith}.

As a simplification we assume that cohesion is zero, as its effect is only noticeable in similar-sized collisions in the diameter range $\lesssim\SI{1}{\kilo\meter}$ \citep{2015PSSJutzi}.  
The model thus assumes fully-damaged material in the solid state, governed by a friction model, given by
\begin{equation}
    Y_\mathrm{d}=\mu_\mathrm{p}p,
\end{equation}
where $\mu_\mathrm{p}$ is the coefficient of friction, a material parameter, and $p$ is pressure. The subscript ``d'' refers to damaged material. Strength does not increase arbitrarily with pressure; it is limited by the yield strength of intact (non-damaged) material,
\begin{equation}
    Y_\mathrm{i}=\frac{\mu_\mathrm{i}p}{1+\mu_\mathrm{i}p/Y_\mathrm{m}},
\end{equation}
where $\mu_\mathrm{i}$ is the coefficient of internal friction and $Y_\mathrm{m}$ is the von Mises plastic limit. The subscript ``i'' refers to intact material. The \emph{full} form of the yield strength is $Y_\mathrm{p}=\min{\left(Y_\mathrm{d},Y_\mathrm{i}\right)}$, which represents a material supported by friction subject to plastic yielding.

We adopt the constitutive model for ``rock materials'' in Table A.1 of \citet{2004MaPSCollins}; here $\mu_\mathrm{p}=0.8$ and $\mu_\mathrm{i}=2.0$ for all simulations. We assume the same coefficients of friction for the iron cores as well; although it is simplistic this choice is unlikely to matter because the pressure at the core-mantle-boundary in all our bodies exceeds $Y_\mathrm{m}$ for iron, \SI{0.68}{\giga\pascal}. The forsterite mantle is subject to friction, when at lower pressure and $Y_\mathrm{m}$=\SI{3.5}{\giga\pascal}. We study the effect of our choice for $\mu_\mathrm{p}$ and $\mu_\mathrm{i}$ in Appendix~\ref{sec:res-solidmodel}.

Yield strength is also temperature-dependent. To capture this effect, we further modulate the yield strength as it approaches the melting point with
\begin{equation}
    Y_\mathrm{T}=Y_\mathrm{P}\tanh\left[\zeta_\mathrm{T}\left(\frac{T_\mathrm{M}}{T}-1\right)\right],
    \label{eq:yield-temp}
\end{equation}
where $\zeta_\mathrm{T}=1.2$ is the thermal softening parameter \citep{1995GRLOhnaka,2004MaPSCollins} and $T_\mathrm{M}$ is the melting temperature. The melting temperature is consistently recovered from the EOS by determining the melting temperature for both materials. This is possible because the \texttt{ANEOS} forsterite EOS \citep{2019AIPStewart} provides the full phase information, while the older EOS for quartz \citep{2007M&PSMelosh} applied in previous work \citep[e.g.,][]{2011PhDReufer,2018IcarusEmsenhuber} does not.

In summary, all simulations are performed with strength of a fully-damaged material modulated by yielding beyond each material's plastic limit, and further by temperature. Strength thus applies to a thin or even unresolved layer in the larger bodies of the database, and to the entire mantle and perhaps some of the core in the smallest bodies in the database. While including strength adds computational cost for planet-scale collisions where strength is not expected to influence outcomes, it ensures constant treatment across the database. Moreover, the simulations involving the most massive bodies are the least computationally expensive, due to the Courant-Friedrichs-Lewy time step criterion being proportional to the spatial resolution (which is itself proportional to the body sizes with a constant number of particles, as in this work). Thus, including friction at the largest scales does not significantly slow down the computation of the database, since most computational effort is on the smaller bodies.

\subsection{Body preparation and initial thermal state}
\label{sec:methods-thermo}

Colliding bodies need to have their initial thermal state specified. This is particularly important for smaller-scale collisions as our SPH model includes friction. The yield strength in the friction model is modulated by the ratio between the temperature of the material undergoing stresses and that of the melting point (see Eq.~(\ref{eq:yield-temp})). 

\begin{figure}
    \centering
    \includegraphics{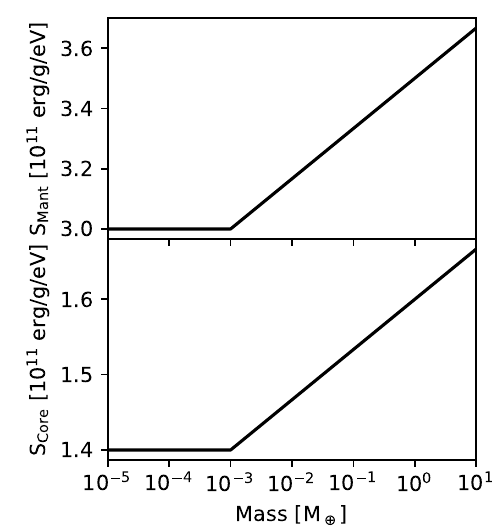}
    \caption{Initial specific entropy for the core ($S_\mathrm{core}$) and mantle ($S_\mathrm{mant}$) as function of the body's mass. The values of entropy are given in the units of \texttt{ANEOS}, \si{\erg\per\gram\per\eV}.}
    \label{fig:entropy}
\end{figure}

\begin{figure*}
    \centering
    \includegraphics{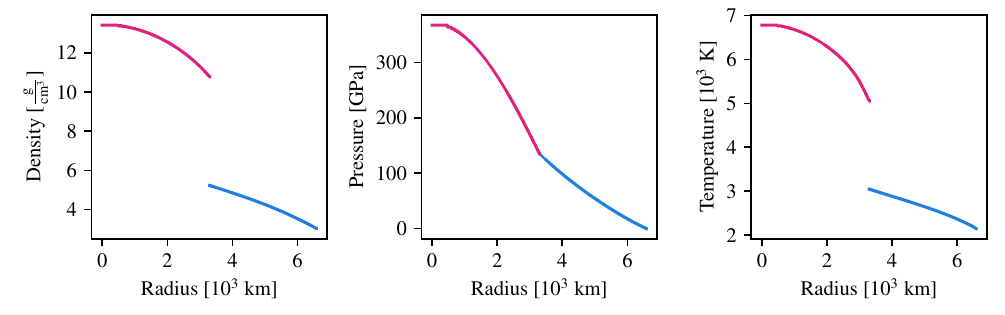}
    \caption{Initial 1D radial profile of a body with mass $M=\SI{1}{\mearth}$ and a core mass fraction $Z=\SI{30}{\percent}$. The red curve corresponds to the iron core and the blue line corresponds to the forsterite mantle.}
    \label{fig:profile-earth}
\end{figure*}

We use isentropic profiles for each of the core and mantle. We leave the initial specific entropy of the core and mantle, $S_\mathrm{core}$ and $S_\mathrm{mant}$ respectively, to be determined. We prescribe thermal profiles such that a large portion of the cores and mantles of the pre-impact bodies are in solid phase, but not far from their melting point. Due to the wide range of body masses that represent our starting conditions, this thermal state is not achievable with a single global value of entropy for the cores or mantles. Hence, we selected the relationship presented in Figure~\ref{fig:entropy}. These have two different regimes, with a constant entropy for all bodies whose masses are below \SI{e-3}{\mearth} and a log-linear relationship above. The selected entropies result in core and mantle temperatures of \SI{1651}{\kelvin} and \SI{1657}{\kelvin} respectively for the bodies at the low-mass end of the range. For larger bodies, a temperature drop occurs at the core-mantle boundary, which is expected. Thermal conduction is not included in our SPH model, so this temperature jump between materials persists through calculations. To highlight the jump in temperature we show a profile on an Earth-like body (mass $M=\SI{1}{\mearth}$ and core mass fraction $Z=\SI{30}{\percent}$) in Figure~\ref{fig:profile-earth}. Note that for the chosen EOS, the outer radius of the body is \SI{6577}{\kilo\meter}, roughly \SI{3}{\percent} larger than the Earth.

The body preparation scheme follows \citet{2021PSJEmsenhuber}: we begin by obtaining a 1D radial profile in hydrostatic equilibrium using the scheme presented in \citet{1991LNPBenz}. The profiles are then mapped onto 3D profiles using the methodology described in \citet{2017MNRASReinhardtStadel} and further relaxed for \SI{6}{\hour} to damp particle motions. To enforce the prescribed initial value, we use a fixed-entropy SPH formulation (rather than evolving energy) during this relaxation step of problem setup.

We choose the resolution so that there are roughly \num{100000} total particles in the simulation (in the target and the impactor). For the smallest mass ratio scenarios ($\gamma=0.05$) this equates to roughly \num{5000} particles in the impactor, since particles are roughly (but not exactly) equal mass in our simulations.
Prior studies \citep[e.g.,][]{2010ChEGAsphaug} have shown that accretion efficiency is resolved to a few percent for SPH simulations with 20,000 and more particles, and \citet{2021MNRASMeier} found that $Q^*_\mathrm{RD}$ (the catastrophic disruption threshold) is converged for \num{e5} particles in SPH simulations using \texttt{ANEOS}, providing additional confidence that our results for lower-energy, hit-and-run scenarios are converged in terms of the largest remnant mass. We also performed simulations with a resolution increased by a factor of ten (roughly 1 million total particles) that show only small differences compared to the nominal resolution (Appendix~\ref{sec:res-res}).

\subsection{Simulation evolution}
\label{sec:methods-evolution}

We set up each collision by placing the bodies at a distance equal to 5 times the sum of the radii, well outside the region where tidal influence begins to deform the bodies (more than twice the Roche limit). The collisions are then evolved for 50 $\tau_\mathrm{coll}$ past initial contact, which is defined as
\begin{equation}
    \tau_\mathrm{coll} = \frac{2\left(r_\mathrm{tar}+r_\mathrm{imp}\right)}{v_\mathrm{coll}}.
\end{equation}
$50\tau_\mathrm{coll}$ corresponds to a bit more than \SI{1}{\day} for the low-velocity collisions (that is, at the mutual escape velocity).
By inspection we find that this duration is sufficient in most cases to determine the collision outcome. However, grazing collisions, where the impactor is captured on a bound orbit with a long duration loopback orbit demand longer integrations. These collisions encompass both graze-and-merge, where the impactor is left on an orbit that intersects with the target, and graze-and-capture regimes. Graze-and-capture refers to collisions where the impactor remains as a bound satellite, akin to the scenario of \citet{2005ScienceCanup} for the origin of Pluto-Charon. 

For cases that end up near the transition between the hit-and-run and graze-and-merge regimes, the simulations and end-states must be analyzed separately \citep{2019ApJEmsenhuberA}. The runner's loopback orbit requires days of physical time, typically, so for returning runners with more than about \SI{\sim10}{\percent} of the impactor's mass, we continue the SPH evolution until it has made at least one more passage of the pericenter. Afterwards, it may be tidally disrupted or partially accreted.
We discriminate two cases in these scenarios. If the orbital period of the runner around the target is smaller than about \SI{7}{\day}, the simulation is further evolved until, usually, no such body remains. We checked that heating due to spurious activation of the artificial viscosity (for instance, due to residual motion inside the bodies) is minimal during the loopback orbit. However, if the orbital period is determined to be longer than \SI{7}{\day}, the simulation is kept in this intermediate state and marked as such. This is to avoid the build up of numerical imprecision in the gravity solver during such an extended period, which could make the final state less realistic. 

\subsection{Simulation analysis}
\label{sec:methods-analysis}

We proceed as in \citet{2019ApJEmsenhuberB} and \citet{2020ApJEmsenhuberA} for analysis of the SPH simulations, and perform additional analysis of body shapes, interior structures, and debris. We begin with a friends-of-friends search of SPH particles to find contiguous bodies. Each of these bodies is replaced by one ``super-particles'' with equivalent mass and momentum for the remaining steps. The second step is to find gravitationally-bound material and identify them as unique clumps. In our work, we consider a minimum clump mass worthy of analysis to be 10 times the mass of the most massive SPH particle, although in practice our results do not depend on this minimum. Particles not part of any clump is considered ``unresolved'' debris. Identified clumps can include co-orbiting pairs of super-particles and multiple-body systems. A last step is to find physical bodies. When a single contiguous body is found in a clump, all remaining particles in such a clump are attributed to that body. When there are more, particles not part of a contiguous body are attributed to the body with which they have the lowest relative energy.

These bodies and clumps provide the basic metrics of collision outcomes, which are archived in machine-readable format. The most massive clump (from the second analysis step) is taken as the largest remnant, whose accretion efficiency $\xi_\mathrm{L}$ is given by Eq.~(\ref{eq:acc_L}). In a perfect merger $\xi_\mathrm{L}$ equals 1 and a value of 0 represents a target with a post-impact mass that is equal to its pre-impact mass. The second most massive clump is defined as the second remnant, for which we define the accretion efficiency of the second remnant of mass $m_\mathrm{S}$:
\begin{equation}
    \xi_\mathrm{S} = \frac{m_\mathrm{S} - m_\mathrm{imp}}{m_\mathrm{imp}} = \frac{m_\mathrm{S}}{m_\mathrm{imp}}-1.
\end{equation}
In a hit-and-run collisions, this second remnant is usually made largely of impactor continuing downrange (i.e., a `runner'). Under most circumstances, mass is eroded from the body, such that $\xi_\mathrm{S}<0$. Material not part of the two most massive remnants is considered debris; either resolved if it is part of a clump or unresolved if not. By analogy to the two largest remnants, debris production is characterised by its accretion efficiency $\xi_\mathrm{D}$, that is its mass in units of impactor masses, or
\begin{equation}
    \xi_\mathrm{D} = \frac{m_\mathrm{D}}{m_\mathrm{imp}}.
\end{equation}
Mass conservation requires that $\xi_\mathrm{L}+\xi_\mathrm{S}+\xi_\mathrm{D}=0$.

We then derive the orbital-dynamical parameters that are sufficient to set up the post-impact bodies on their new orbits following an identified collision in an \textit{N}-body simulation. In this step, we use the radius of an equivalent body with the same mass and core mass fraction obtained from 1D calculations with initial thermodynamic profile, using the results shown in Appendix~\ref{sec:1d}. This is to use the same radii as in the \textit{N}-body. Using the post-collision radii would either create additional complexity or lead to the use of inconsistent radii. Having consistent radii is important to ensure that the relative velocity and impact parameter are correct.

For collisions with no second remnant, the velocity change (post-impact orbit) of the target is computed, where to conserve momentum in the center-of-mass frame, it has the equal but opposite momentum of the debris.
For collisions with two major remnants after the initial collision, whether they end up being hit-and-run or graze-and-merge, we determine the final (outgoing) orbits of both the bodies. Since we do not include pre-impact rotation, we can assume the outgoing orbits are in the same plane as the incoming colliding bodies. The true anomaly is not determined or considered herein, as the initial location of the remnants after the collision will be prescribed according to the capabilities and design of \textit{N}-body codes that will use our results and simulate the evolution of the remnants. 

As in \citet{2020ApJEmsenhuberA}, there are three additional orbital-dynamical parameters from the post-impact remnants that need to be computed to describe their orbits in sufficient detail for N-body simulations.
We compute the orbital energy as
\begin{equation}
\epsilon'=\frac{{v'}^2_\mathrm{coll}}{{v'}^2_\mathrm{esc}}-1=\frac{{v'}^2_\infty}{{v'}^2_\mathrm{esc}},
\end{equation}
where the primed ($'$) quantities are computed from the properties of the largest and second remnants instead of the target and impactor, respectively. The second equality is only valid in the case of an unbound orbit.
Next we compute the impact parameter
\begin{equation}
b'=\frac{h'}{R_\mathrm{coll}'v_\mathrm{coll}'}
\end{equation}
where $h'$ is the norm of the specific relative angular momentum vector. 
Finally we have the shift of the orbit's pericenter
\begin{equation}
\Delta\varpi=\varpi'-\varpi.
\end{equation}
Together, these orbital-dynamical parameters are sufficient to set up the post-impact orbits in an N-body simulation.

\section{SPH Results}
\label{sec:results}

While there is a great wealth of information in this new, publicly-available database to be leveraged, we present a few major conclusions that are worth highlighting. In particular, we find results related to remnant shape and the nature of debris as a function of total mass (reported in Table~\ref{table:sph}), which provide new insights into planet formation physics.

\subsection{General SPH outcomes}
\label{sec:res-gen}

\begin{table*}
    \centering
    \caption{Description of the table headers for the outcomes of SPH simulations. The first eight columns of the table are the initial conditions while the rest summarize the outcomes of the simulations.}
    \label{table:sph}
    \begin{tabular}{c|ccl}
        \hline
        Group & Quantity & Unit & Description \\
        \hline
        \hline
        & set & & Name of simulation set \\
        & num & & Simulation number in the set \\
        \hline
        \multirow{6}*{Collision Parameters} & $m_\mathrm{tar}$ & [\si{\mearth}] & Mass of the target \\
        & $\gamma$ & & Impactor-to-target mass ratio \\
        & $Z_\mathrm{tar}$ & & Target's core mass fraction \\
        & $Z_\mathrm{imp}$ & & Impactor's core mass fraction \\
        & $\frac{v_\mathrm{coll}}{v_\mathrm{esc}}$ & & Relative velocity at initial contact \\
        & $\theta_\mathrm{coll}$ & [deg] & Impact angle \\
        \hline
        & Regime & & Automated classification (see main text) \\
        \hline
        \multirow{13}*{End state} & $\xi_\mathrm{L}$ & & Accretion efficiency of largest remnant \\
        & $\xi_\mathrm{S}$ & & Accretion efficiency of second remnant \\
        & $\xi_\mathrm{D}$ & & Accretion efficiency of debris \\
        & $\Omega_\mathrm{L}$ & [rad/hr] & Spin rate of the largest remnant \\
        & $Z_\mathrm{L}$ & & Largest remnant's core mass fraction \\
        & $I_\mathrm{Z,L}$ & & Largest remnant's moment of inertia factor along spin axis \\
        & $I_\mathrm{A,L}$ & & Largest remnant's volume-averaged moment of inertia factor \\
        & $\xi_\mathrm{P}$ & & Accretion efficiency of primary physical body \\
        & $\gamma_\mathrm{P}$ & & Mass ratio of the second to primary physical body \\
        & $n_\mathrm{P}$ & [rad/hr] & Mean motion of the second physical body's orbit around the primary \\
        & $N_\mathrm{res}$ & & Number of resolved bodies \\
        & $\frac{v_\mathrm{RMS}}{v_\mathrm{esc}}$ & & RMS of the velocity of debris \\
        & $\frac{v_\mathrm{RMS,\infty}}{v_\mathrm{esc}}$ & & RMS of the velocity of debris at infinity \\
        \hline
        \multirow{15}*{After 1\textsuperscript{st} enc.} & $\xi_\mathrm{L}$ & & Accretion efficiency of largest remnant \\
        & $\xi_\mathrm{S}$ & & Accretion efficiency of second remnant \\
        & $\xi_\mathrm{D}$ & & Accretion efficiency of debris \\
        & $\Omega_\mathrm{L}$ & [rad/hr] & Spin rate of the largest remnant \\
        & $Z_\mathrm{L}$ & & Largest remnant's core mass fraction \\
        & $I_\mathrm{Z,L}$ & & Largest remnant's moment of inertia factor along spin axis \\
        & $I_\mathrm{A,L}$ & & Largest remnant's volume-averaged moment of inertia factor \\
        & $\epsilon'$ & & Scaled orbital energy of the second remnant's orbit \\
        & $b'$ & & Impact parameter of the second remnant's orbit \\
        & $\Delta\varpi$ & [rad] & Shift of pericenter of the second remnant's orbit \\
        & $\Omega_\mathrm{S}$ & [rad/hr] & Spin rate of the second remnant \\
        & $Z_\mathrm{S}$ & & Second remnant's core mass fraction \\
        & $I_\mathrm{Z,S}$ & & Second remnant's moment of inertia factor along spin axis \\
        & $I_\mathrm{A,S}$ & & Second remnant's volume-averaged moment of inertia factor \\
        & $N_\mathrm{res}$ & & Number of resolved bodies \\
        & $\frac{v_\mathrm{RMS}}{v_\mathrm{esc}}$ & & RMS of the velocity of debris \\
        \hline
    \end{tabular}
    \tablecomments{This table is available in its entirety in machine-readable form as supplementary online material.}
\end{table*}

The ``regime'' column in Table~\ref{table:sph} is a flag that is automatically set from the analysis of a collision. It determines which properties of the simulation are returned. It is based on the number of significant remnants that are found, where a significant remnant is taken to be a body whose mass is at least \SI{10}{\percent} that of the impactor; considering the resolution of the simulations and the range of impactor-to-target mass ratios covered, this means that the significant renmants classification is restricted to bodies of at least \num{\sim500} SPH particles. The core mass fractions are only determined on significant remnants to avoid computing statistics on bodies that have low numbers of SPH particles in the simulation and thus may be under-resolved. 
To accommodate graze-and-merge collisions, we provide two sets of outcomes for the simulations results: the first for the end state of the simulation and the other after only a single encounter. Depending on the type of collision, these may or may not be at the same time.

The different regime labels are:
\begin{enumerate}
    \item (727 entries or \SI{58.2}{\percent}) A hit-and-run collision where two unbound significant remnants exist. Per our definition of significant remnant, this includes erosive hit-and-runs. Here all the values are computed at the end of the simulation.
    \item (218 entries or \SI{17.4}{\percent}) A finished graze-and-merge collision, where two significant remnants were temporarily gravitationally bound after first contact, and collided again thereafter in an accretion. Here all the properties are determined first at apocenter and then at the end of the simulation.
    \item (10 entries or \SI{0.8}{\percent}) An unfinished graze-and-merge collisions, where the loopback orbital period is too long to evolve back to apocenter using SPH. Here only the properties after the initial grazing encounter are determined, from the end state of the simulation.
    \item (281 entries or \SI{22.5}{\percent}) A collision that results in a single major remnant. Here the accretion efficiency and core mass fraction are determined only for the largest remnant, at the end of the simulation.
    \item (14 entries or \SI{1.1}{\percent}) A fully erosive event where no significant remnant exists. Only the accretion efficiencies (negative) are determined in this case, at the end of the simulation, as there is insufficient resolution to determine the other properties.
\end{enumerate}

We observe that hit-and-runs are the most represented collisions in our database. This is consistent with the fact that these collisions are found for a wide range of impact angles for our favored impact velocities \citep{2019ApJCambioni}. Only 14 collisions (or \SI{1.1}{\percent} of the total) are super-catastrophic. This is unsurprising as they only occur at large velocities and low impact angles \citep{2012ApJLeinhardt}, which our selected velocity distribution does not favor.

For unfinished grazing collisions---those labeled \texttt{3}---we refrain from providing the final values. The final outcome is less relevant in the context of planetary formation, where the loopback orbit would be perturbed by third bodies, such as the Sun \citep{2019ApJEmsenhuberB}.

Due to the sensitivity on impact angle and velocity, grazing collisions can have a diversity of specific outcomes even for similar initial conditions. Example end states include the capture of a substantial part of the runner as a satellite, or producing bodies that are far from their original hydrostatic equilibrium. We provide a selection of these outcomes in Figure~\ref{fig:scatter} that will be discussed in more detail in the following sections.

\subsection{Capture as satellite}
\label{sec:res-sat}

\begin{figure*}
    \centering
    \makebox[2.3in]{\Large Low-mass capture}
    \makebox[2.3in]{\Large High-mass capture}
    \makebox[2.3in]{\Large Unusual shape}
    \includegraphics[width=2.3in]{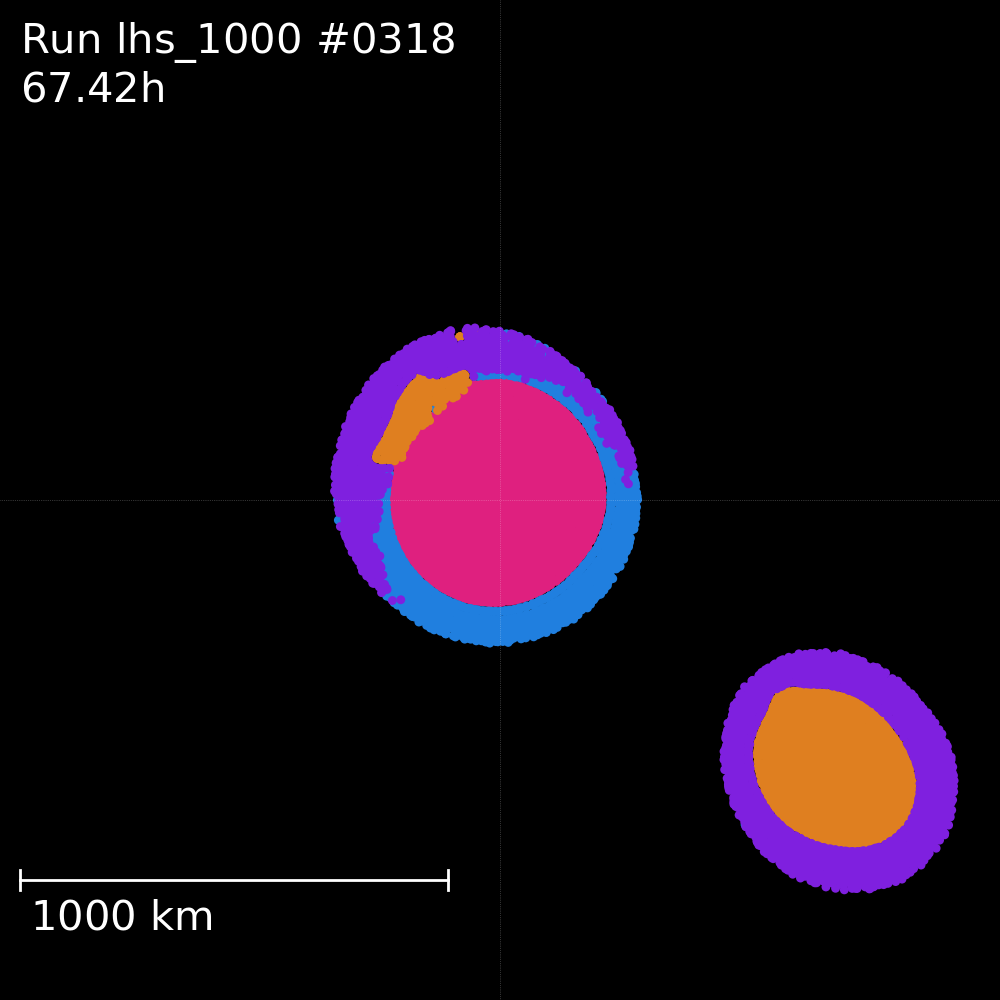}
    \includegraphics[width=2.3in]{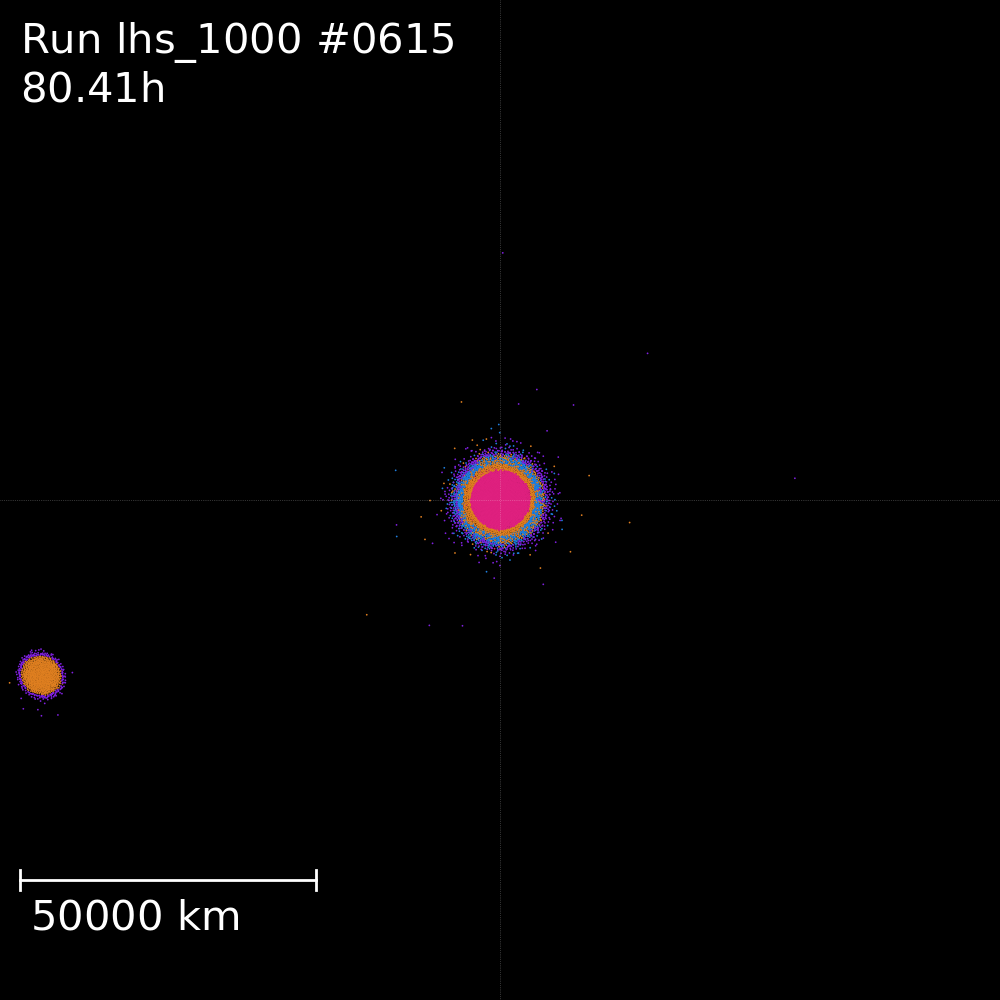}
    \includegraphics[width=2.3in]{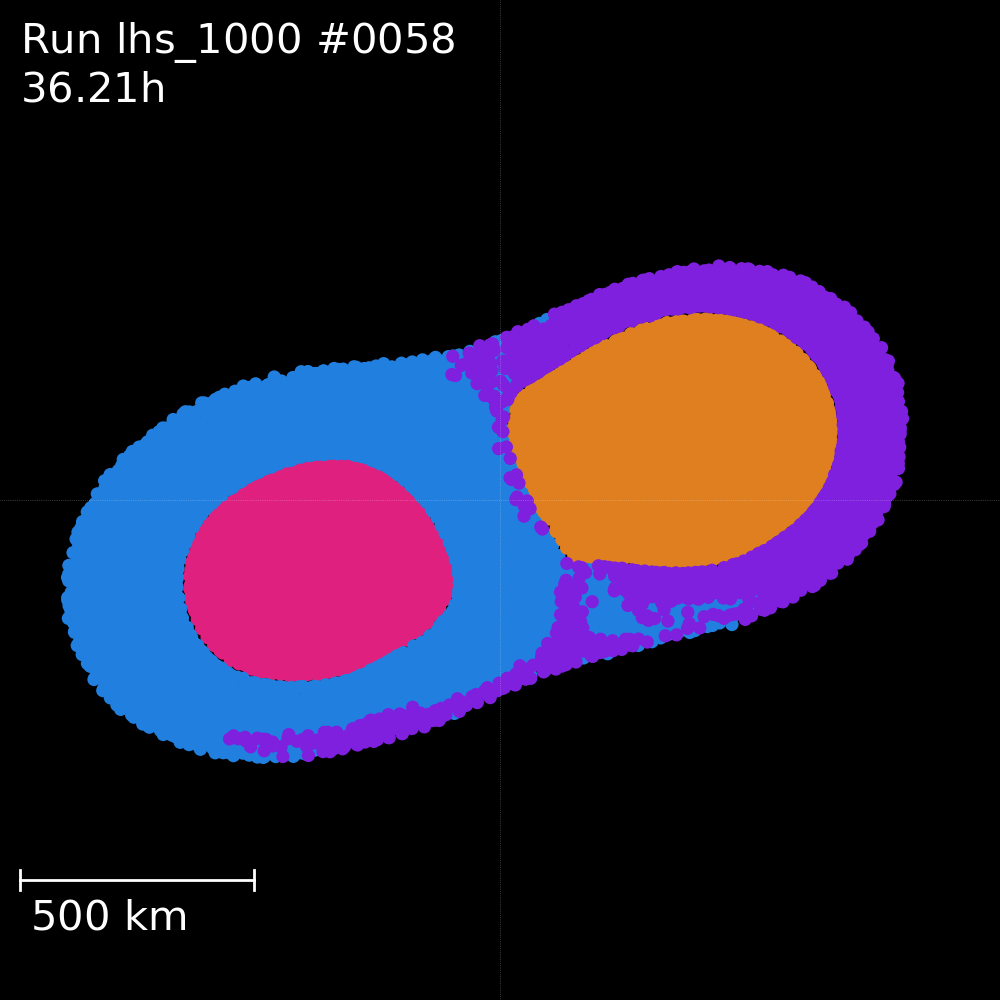}
    \includegraphics{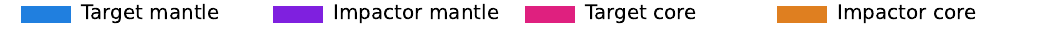}
    \caption{Final state of a selection of three, low-velocity grazing simulations, each with different masses of the colliding bodies (left: $m_{\rm{tot}}\approx~1.9\times10^{-4}~M_{\oplus}$, middle: $m_{\rm{tot}}\approx~2.2~M_{\oplus}$, right: $m_{\rm{tot}}\approx~4.0\times10^{-4}~M_{\oplus}$); note the scale bars. Left and center panels show capture as satellite; the right panel shows unusual body shape. Colors represent material and origin body, as indicated in the legend: blue for target's mantle, purple for impactor's mantle, red for target's iron core, and yellow for impactor's core.}
    \label{fig:scatter}
\end{figure*}

\begin{figure}
    \centering
    \includegraphics{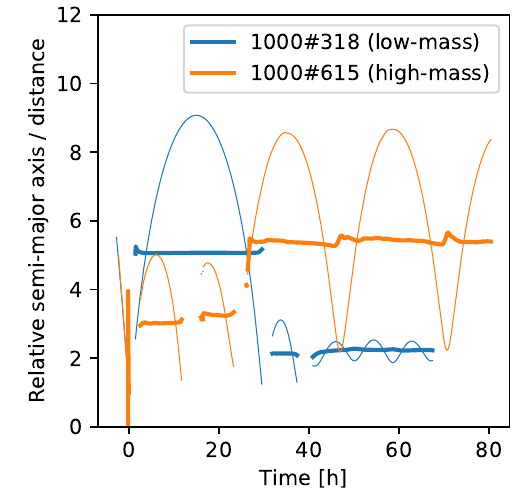}
    \caption{Time evolution of the relative semi-major axis (thick lines) and distance (thin lines) of the two largest remnants for grazing collisions resulting in satellite capture shown in Figure~\ref{fig:scatter} (left and central panels). The values are given in terms of the sum of the radii of equivalent bodies with the given mass and core mass fraction, according to the results from Appendix~\ref{sec:1d}. Gaps indicate physical contacts between the two bodies. The satellite resulting from the collision between the low-mass bodies remains on a low-eccentricity orbit close to the primary, while the satellite in the high-mass collision is left on an eccentric orbit further out.}
    \label{fig:satellite-dist}
\end{figure}

\begin{figure}
    \centering
    \includegraphics{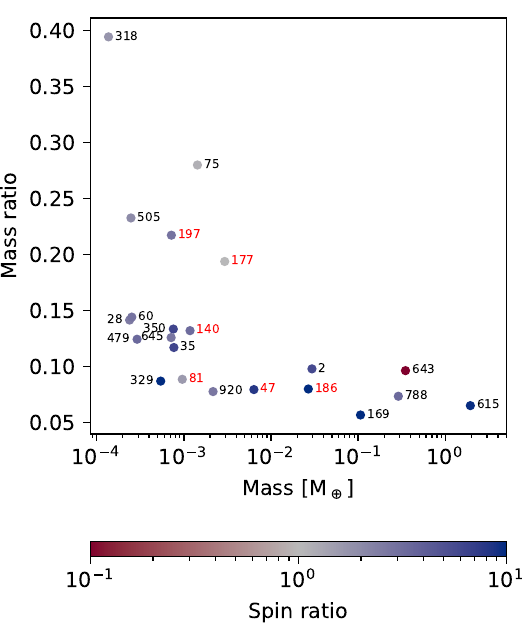}
    \caption{Mass ratio of the satellite-to-primary pairs ($\gamma_\mathrm{P}$) as function of the primary mass (computed as $m_\mathrm{tar}\left(1+\gamma\xi_\mathrm{P}\right)$), for all simulations ending in co-orbiting secondary remnants. The number in the plots refer to the corresponding run numbers: black for those in the 1000-simulation set and red for those in the 250-simulation one. The color scale represents the ratio between the orbital frequency of the satellite and the spin rate of the primary ($n_\mathrm{P}/\Omega_\mathrm{L}$), which is 1 for synchronous orbit and must be $<1$ for stable, outward tidal migration.}
    \label{fig:satellite-list}
\end{figure}

One of the peculiar outcomes of grazing collisions are situations where most of the impactor (the runner) ends up forming a satellite orbiting the largest remnant. This is akin to a proposed scenario for the formation of the Pluto-Charon binary \citep{2005ScienceCanup} and some scenarios for the formation of our Moon \citep{1987IcarusBenz}. We remind the reader that a bound satellite---like any bound material---is considered part of the largest remnant in this study.

We also examine the long-term evolution of the satellite. 
If the satellite orbit is below the corotation radius, tidal friction will transfer angular momentum over time from the satellite's orbit to the spin of the central body, and it will spiral down on a timescale dictated by tidal dissipation and reimpact the central body. This results in the accretion of the majority of the satellite's mass and ejection of a fraction of the mass to release angular momentum. If the satellite is captured outside the corotation radius then the opposite occurs, and angular momentum is transferred to the satellite's orbit and it migrates further out.

Examples of satellite capture in our database are shown in Figure~\ref{fig:scatter}, left and center panels. One scenario is a collision with $m_\mathrm{tar}=\SI{1.154e-4}{\mearth}$ and $\gamma=\num{0.657}$, and whereas the other is a higher-mass target $m_\mathrm{tar}=\SI{1.618}{\mearth}$ with $\gamma=\num{0.376}$. Both occur near the mutual escape velocity($v_\mathrm{coll}/v_\mathrm{esc}=\num{1.02}$ and \num{1.00}, respectively) at grazing incidence (\SI{75.1}{\degree} and \SI{79.5}{\degree}, respectively) which is characteristic of satellite capture scenarios.

Figure~\ref{fig:satellite-dist} shows how multiple loopback encounters can lead to satellite capture. The blue thin curve plots the relative distance between the largest remnants versus time for the low-mass case in Figure~\ref{fig:scatter} (left panel). The orange curve shows the same information for the high-mass case (center panel). The gaps in the curves indicate the moment the bodies are in physical contact. An ``arm'' (continuous link of material) usually connects the two bodies until the bodies reach a distance up to several mutual radii. During this time, the friend-of-friend algorithm cannot identify the central body and satellite as separate objects and the orbit cannot be determined. The results show that the pathway toward capture differs between the two collisions.

For the low-mass case each close encounter leads to dissipation in the satellite which decreases the apocenter. Little orbital angular momentum is transferred onto the target, likely due to the inclusion of material strength in our simulations, which limits its deformation. Consequently, the orbit becomes more circular. For the high-mass case, both the satellite and target are heavily deformed during each loopback encounter, and much of the impactor's material is transferred onto the target during successive encounters. This raises the pericenter when tidal deformation induces a torque on the remnant of the impactor, which increases its angular momentum.

In the high-mass scenario in Figure~\ref{fig:satellite-dist} (orange lines), it is the third encounter at \SI{\sim23}{\hour} that is the most effective at establishing a relatively stable orbit for the impactor.
The satellite's orbit at the end of our simulation is eccentric (see difference between thin curve and thick curve, representing radius and semi-major axis of the orbit respectively), resulting in the satellite spending most of the time many radii away from the target. On its subsequent close approaches, tidal destruction is avoided due to the sufficiently high pericenter. We evolve these simulations long enough for the satellite to make multiple pericenter passages to check that satellites are not destroyed during this phase.

For all capture scenarios, we report the mass ratio of the orbiting pairs in Figure~\ref{fig:satellite-list}. For this analysis, we ensure that there is at least one pericenter passage of the loopback orbit. We find that satellites with more than \SI{\sim10}{\percent} of the mass of the primary occur around primaries with masses below \SI{2e-3}{\mearth}. Friction likely aids in the capture of such massive moons, as exemplified by the low-mass capture scenario shown in Figures~\ref{fig:scatter} and~\ref{fig:satellite-dist} where the satellite remains at about twice the mutual radii. Without friction, such a satellite would likely be disrupted by tidal forces, as it lies inside the Roche limit of a fluid body.
Satellites are obtained around more massive bodies, but are of a small mass ratio with respect to the post-impact target. This outcome is consistent with prior work \citep{2022NatCoNakajima} on the limited ability of larger planets to form fractionally massive moons; however, the result here is for collisional capture and is a direct aspect of the giant impact phase itself. 
Also, as a general rule, fractionally massive surviving clumps are less common at larger scales of giant impacts because of the greater relative energies involved \citep{2014NatGeoAsphaug}.

\subsection{Final body shapes}
\label{sec:res-shape}

\begin{figure}
	\centering
	\includegraphics{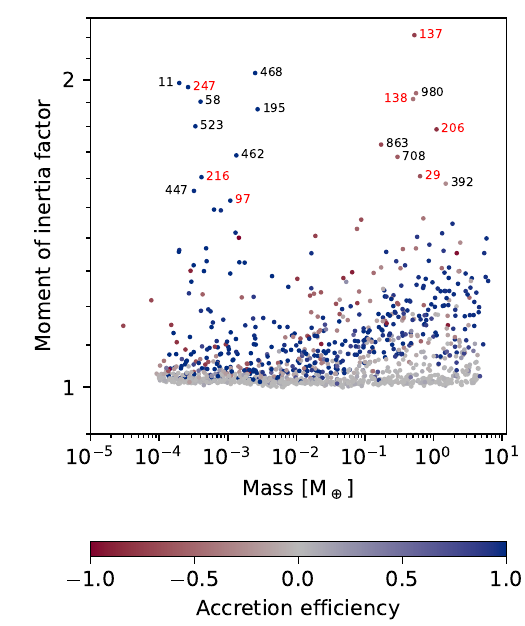}
	\caption{The moment of inertia factor of the post-impact largest remnant (y axis) as a function of its mass, with accretion efficiency of the largest remnant ($\xi_\mathrm{L}$) in color. The moment of inertia factor is the moment of inertia divided by that of a hydrostatic body with the same mass and core mass fraction, as outlined in Appendix~\ref{sec:1d}. Both axes use logarithmic scales. To avoid catastrophic events, only collisions where $\xi_\mathrm{L}>-1$ are shown. The simulation number is shown when the moment of inertia factor is larger than 1.5. A black label is used for runs in the 1000-simulation set and red for those in the 250-simulation set.}
	\label{fig:inf-lr}
\end{figure}

An outcome of low-mass, graze-and-merge collisions is the emergence of non-spherical bodies. Friction plays a role in sustaining shapes that are far from hydrostatic equilibrium \citep[e.g.,][]{2009ApJTanga,2018AASugiura,2019IcarusJutzi}. For example, in Mars-sized planets and smaller, friction forces can counteract gravity, allowing for the stranding of impact cores in mantle \citep{2018IcarusEmsenhuber}. 
On geologic timescales, these non-hydrostatic shapes and irregular core structures might relax, lowering the moment of inertia and changing the spin state.

An example is shown in the third panel of Figure~\ref{fig:scatter}, which shows a cross-section of simulation \texttt{58} of the 1000-simulation set. These scenarios are analogous to hypotheses for the formation of contact binaries such as 486958 Arrokoth \citep{2019ScienceStern} and 67P Churyumov-Gerasimenko \citep{2015ScienceJutziAsphaug,2017AAJutziBenz}, although those scenarios invoke crushable solid bodies while our models are at much larger scales and consider bodies that start out with metallic cores. Here it can be seen that the two bodies only slightly mix and the cores remain separated.

The non-spheroidal shape of the final body, in these extreme cases, is not only maintained by material strength, but also from the spin induced by the collision. The production of non-spherical shapes generally require grazing mergers at velocities close to $v_\mathrm{esc}$. In these cases, the two cores do not immediately intersect during the initial collision, yet there is enough dissipation in the initial encounter that the bodies remain bound, rotating quickly and maintaining a spin-supported shape. A caveat of our analysis, however, is that bodies are not spinning before the collision, so pre-impact spin is not taken into account.

We detected bodies with non-spherical shapes by computing the moment of inertia of all massive remnants and comparing the values that of an equivalent nonrotating spherical body of the same composition under hydrostatic equilibrium. Figure~\ref{fig:inf-lr} is a scatter plot of the ratio between these values as a function of mass.  Ratios close to 1 indicate bodies whose internal structure is comparable to that of initial bodies.

The analysis reveals several features. Remnants from hit-and-run collisions, where $\xi_\mathrm{L}\approx 0$ (in gray in the figure), in the end have their global structures barely affected by the collision. Accretionary or erosive collision usually lead to increased moment of inertia, due to spin-induced deformation (equatorial bulge), for accretionary events, or pressure release, for erosive events. Finally, we see two regions that exhibit moment of inertia factors larger than two: one for small bodies (below \SI{\sim2e-3}{\mearth}) and another for large bodies (around \SI{1}{\mearth}). 

For small bodies below \SI{\sim2e-3}{\mearth} (or \SI{2000}{\kilo\meter} diameter) for the modeled terrestrial compositions, the high moment of inertia factor reflects the peculiar shapes. In Figure~\ref{fig:scatter} (right panel) the contact binary would plot towards the top left of Figure~\ref{fig:inf-lr}. It is not a unique case in our database; however, this compound core scenario only happens for small bodies. The specific compound-core outcome and other aspects of shape depend on the influence of friction, especially in the silicate material that supports the denser core material, where for consistency with past and ongoing research we have adopted the coefficients of friction in Table A1 of \citet{2004MaPSCollins}.
The upper limit of \SI{\sim2e-3}{\mearth}, which coincides with the limit for satellites with large mass ratios (Sect.~\ref{sec:res-sat}), suggests that material strength has global-scale effects up to that mass. We note that this upper limit depends on the constitutive strength model and its parameters. As a consequence, bodies with significantly higher initial temperature or lower von Mises plastic limit $Y_\mathrm{m}$ would exhibit the transition to more spherical shapes at smaller sizes.

In the case of the accretionary events in Figure~\ref{fig:inf-lr}, the increased moment of inertia factor is due to a combination of two effects. The predominant one is the spin, which causes a rotational bulge. Unlike the peculiar shapes we just discussed, however, the shape of these bodies is nearly symmetric about their spin axis. Other effects, limited to the masses above \SI{\sim e-2}{\mearth} are thermal in nature. This is because mergers, especially those between similar-mass bodies (that is, for $\gamma\approx1$), release the most binding energy \citep{2013IcarusAsphaug,2020JGRECarter}, increasing temperature. These bodies will also preferentially form distended vapor-rich outer structures due to shock heating in the largest remnant \citep{2017JGRELockStewart}. This in turn expands the bodies relative to a starting configuration \citep{2017JGRELockStewart} and increases their moment of inertia.

Strongly erosive collisions ($\xi_\mathrm{L}\approx-1$) also exhibit similar expanded states because of pressure release, as in these cases usually only the most central part of the body remains. Even-lower values of $\xi_\mathrm{L}$ are obtained in super-catastrophic disruptions, and show the same effect, but they are not plotted because bodies are comprised of relatively few particles. 

\subsection{Debris properties}
\label{sec:res-debris}

One of the main objectives of our new simulations is to better characterize the debris to represent it with known accuracy in planet formation studies. For this we need to understand not only debris production efficiency $\xi_D$ but also its velocity and the size distribution.

\subsubsection{Fragment sizes and number}

\begin{figure}
    \centering
    \includegraphics{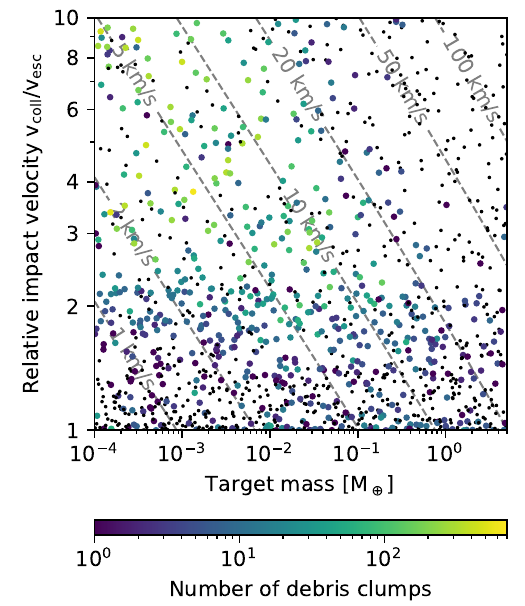}
    \caption{Scatter plots of the number of resolved (\num{>10} SPH particle) fragments, excluding the largest and second remnants, found in our simulation database as function of target mass and relative velocity $v_\mathrm{coll}/v_\mathrm{esc}$. The color represents number of fragments in a single simulation. Collisions not producing any fragment are shown by smaller black circles. Although these points include simulations with various impactor-to-target mass ratios, we show absolute impact velocities for the target mass alone, assuming a core mass fraction of \SI{30}{\percent}). It can be seen that the number of resolved clumps decreases with increasing target mass.}
    \label{fig:debris-num}
\end{figure}

Resolvable fragments are clumps that are obtained in the debris field, identified in the same way as we identify the largest and second remnants, by performing friend-of-friend and gravity searches in that order. The smallest mass that can be resolved depends on the numerical resolution of the simulations, however the size-frequency distribution as a whole is much less sensitive to numerical resolution (Appendix~\ref{sec:res-ressfd}). We assume that a fragment requires at least ten particles to be meaningful, for our database with roughly \num{e5} particles per simulation, and thus only clumps larger than \num{\sim e-4} the total mass are regarded as significant. The particles in the smallest clumps have less than a full neighbor-list (approximately 50 particles), and while this means that SPH forces are not computed accurately, the self-gravity that causes clumping is accurate. We have therefore verified that the results discussed here remain if we change the fragment resolution limit from \num{10} to \num{50} to \num{100} SPH particles.

Fewer than half of our modeled collisions produce fragments 10 SPH particles in size or larger, besides the two largest remnants. No debris fragment is larger than \SI{10}{\percent} of the total mass and only few collisions produce any fragment larger than \SI{1}{\percent} of the total mass. The only exception is the second remnant, which we do not include in the debris size-frequency distribution. Most massive fragments are found in either graze-and-merge or disruptive hit-and-run collisions, where the impactor is broken in multiple fragments. 

It is extremely difficult for \textit{N}-body simulations to track debris directly, as the number of bodies (\textit{N}) would increase dramatically whenever there is a collision. Thus, debris need to be treated differently from the two largest remnants, with a more statistical approach, such as tracer particles \citep[e.g.,][]{2012AJLevison,2019IcarusWalshLevison}, bodies of a fixed mass \citep[e.g.,][]{2013IcarusChambers}, or as a background surface density \citep[e.g.,][]{2015ApJCarter}. Possible statistical approaches include looking for correlations between debris size and velocity (Sect.~\ref{sec:res-debdyn}) or fitting a size-frequency distribution to the bodies in the debris field (as in Appendix~\ref{sec:res-res}). Both require that at least a few clumps are reliably determined, and we thus investigated the conditions that are favorable for the production of many resolved debris clumps. We find that a well-populated size distribution of debris is generated in graze-and-merge and shallow angle, erosive collisions. Graze-and-merge ejects clumps due to the large angular momentum involved \citep{2013IcarusAsphaug} and shallow-angle erosive collisions ($v_{\rm{imp}}/v_{\rm{esc}}\geq2$) naturally produce ample amounts of debris.

We further find that low-mass bodies ($m_\mathrm{tar}\lesssim\SI{0.1}{\mearth}$) are favored for the production of a large number of resolved debris clumps. We illustrate this in Figure~\ref{fig:debris-num}, where the number of fragments for each collision is shown as functions of the target's mass $m_\mathrm{tar}$ and the relative collision velocity $v_\mathrm{coll}/v_\mathrm{esc}$. Here we observe that there is a strong transition at $v_\mathrm{coll}\approx\SI{20}{\kilo\meter\per\second}$, which corresponds to $m_\mathrm{tar}=\SI{0.1}{\mearth}$ for $v_\mathrm{coll}/v_\mathrm{esc}\approx 4$. While the maximum number of fragments identified in any collision is larger than \num{400}, none of the collisions above \SI{20}{\kilo\meter\per\second} has more than \num{40} resolvable fragments. At even larger velocities ($v_\mathrm{coll}\approx\SI{50}{\kilo\meter\per\second}$) only few collisions report any resolvable fragment; nearly all the debris are found in unresolved particles. This effect could be due to stronger shocks or vaporization being more common at larger absolute velocity, both of which can prevent the formation of debris clumps.

Our debris results should be taken with caution. First, we use a constant number of particles for our simulations, which means that the minimum mass of resolved fragments scales with the target mass. Therefore fragments of a given absolute size may be resolved in collisions involving low-mass bodies and not with high-mass bodies. Second, high-velocity collisions involve extreme amounts of heating, where numerical effects in SPH may be exacerbated.

\subsubsection{Debris dynamics}
\label{sec:res-debdyn}

\begin{figure*}
    \centering
    \includegraphics{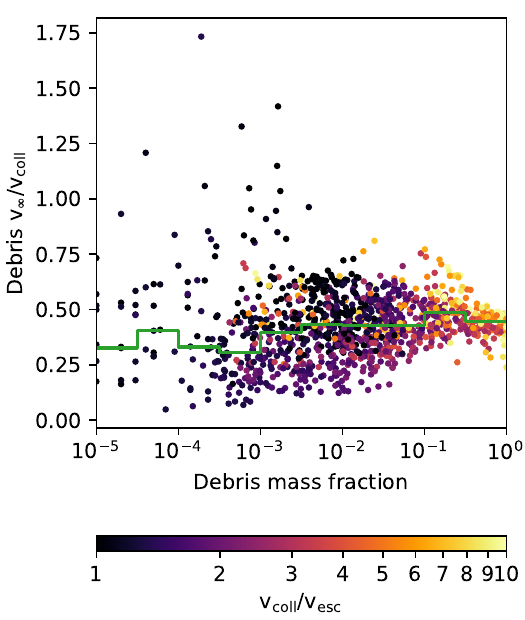}
    \includegraphics{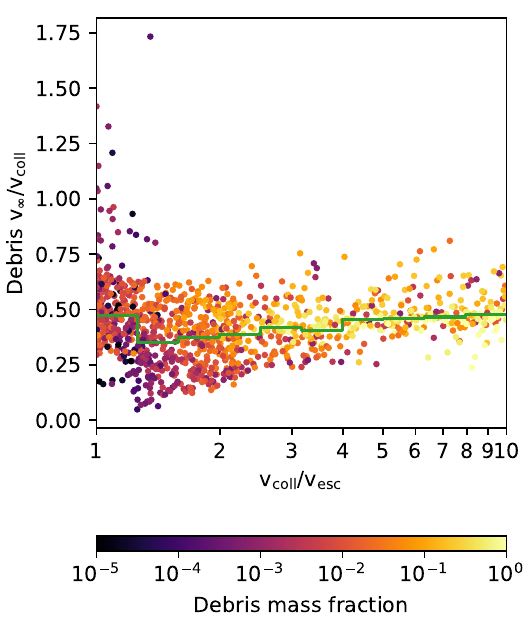}
    \caption{Debris velocity at infinity given in terms of the relative velocity between the initial bodies (target and impactor) at initial contact for all the simulations of this study. Left panel shows the values as function of the mass fraction of debris at the end and color coded by the initial $v_\mathrm{coll}/v_\mathrm{esc}$. Right panel shows the same data with the horizontal axis and color code inverted. The step-wise line shows the median of the values in the given range.}
    \label{fig:debris-vel}
\end{figure*}

Collision after collision, the fraction of the total mass that becomes debris may become significant with time \citep{2020ApJEmsenhuberA}. These debris are reaccreted by the large bodies on timescales that be millions of years \citep{2012MNRASJacksonWyatt}, although some works suggest that the mass fraction is never significant at any given time \citep{2015ApJCarter}. As such, debris orbital evolution must be modeled consistently during \textit{N}-body calculations. For this purpose we report post-impact velocities and their correlation with remnant sizes.

First, we compute the root-mean-square (RMS) of the debris velocity distribution relative to the center of mass, reported as $v_\mathrm{RMS}$ in Table~\ref{table:sph}. This value is computed from their specific kinetic energy. Because the debris are in the gravitational potential of the largest remnants, this value is time-dependent. To account for this, we also report the velocity of the debris calculated using the total orbital energy (kinetic and potential) of the debris field ($v_\mathrm{RMS,\infty}$).

We then check for any correlation between debris size and their velocity using the Spearman's rank correlation coefficient. We find that over the \num{1250} simulations presented here, \num{591} have a sufficient number of resolved debris such that a correlation can be computed. Of those, \num{62} have \textit{p-values} consistent with there being no correlation lower than \SI{1}{\percent}, indicating that the correlation is statistically significant. Of those, \num{10} have a positive correlation with debris size and velocity and \num{52} have a negative correlation. Thus, only a minority (about \SI{10}{\percent}) of our simulated collisions with resolved debris fields show correlations between debris size and velocity. In sum, we find that debris velocity and mass are not correlated in a systematic way.

To broadly describe debris velocity fields in our all simulations, we compute the mean velocity of the debris at infinity in terms of the relative velocity of the precursor bodies at initial contact (see Figure~\ref{fig:debris-vel}).
The left panel shows that collisions producing the most debris (shown towards the right of the panel) have velocities that trend near a value of about \num{0.45}, with a significant dispersion around the mean due to the other parameters (mainly the mass ratio $\gamma$). Collisions that produce a smaller fraction of debris exhibit the greatest dispersion and a lower mean velocity; however, capturing debris velocity accurately is less important for those events, from the point of view of planet formation, as their debris constitutes a much smaller fraction of the total mass. 
The right panel of Figure~\ref{fig:debris-vel} shows that the debris velocity in terms of $v_\mathrm{coll}$ is largely independent of $v_\mathrm{coll}/v_\mathrm{esc}$ of the incoming bodies. This indicates that $v_\mathrm{coll}$ is a better reference for the debris velocity rather than, for instance, $v_\mathrm{esc}$.

As a rule of thumb, we find that the mean debris velocity at infinity $v_\mathrm{RMS,\infty}$ is about half that of the collision velocity $v_\mathrm{coll}$. This can be significantly greater than $v_\mathrm{esc}$ of the largest remnant. Thus, the assumption that the debris always leave at near-escape velocities, which is made by \citet{2013IcarusChambers} and other works based on their prescription \citep[e.g.,][]{2019IcarusClement}, severely underestimates debris velocity. The consequence is an artificial damping of the velocities of a late-stage accreting planetary system and, as a potential consequence, an artificial limitation to the exchanges of materials between formation regions. The debris velocities we obtain differ significantly from the ejecta velocity distributions of collisions reported in \citet{2012ApJLeinhardt}; this is understandable for two reasons. First, their results rely on a discrete element model (DEM) instead of solving the Euler equations, and thus are idealized representations of the collisional physics at the high velocities of planet formation. Second, and more fundamentally, their results are computed for the catastrophic disruption regime (their Figure 5), in which the escaping velocities are governed by impact ejection mechanics rather than gravitational accretion dynamics. Our focus is on the regimes relevant to planetary accretion, so our database barely extends into their regime of catastrophically disruptive collisions (Table~\ref{tab:vel-dist}). Debris are not blasted from the target so much as they are flung out by the complex gravitational interactions of two similar-mass bodies, which explains why the debris velocities in our database are governed by $v_\mathrm{esc}$.

If the relative velocity at infinity $v_\mathrm{RMS,\infty}$ is half of $v_\mathrm{coll}$, then for collisions where $v_\mathrm{coll}/v_\mathrm{esc}<2\sqrt{3}/3\simeq 1.15$, debris velocity at infinity $v_\mathrm{RMS,\infty}$ can be larger than the relative velocity of the incoming bodies at infinity $v_\infty$. This is not paradoxical, as it happens in the case of strong gravitational torques by the more massive bodies in a close approach, and the subsequent ejection of arms of material. Collisions with such low impact velocities usually produce relatively little debris in any case, as can be seen by the color scale on the right panel of Figure~\ref{fig:debris-vel}.

Several simulations exhibit debris with negative total energy and are not represented in Figure~\ref{fig:debris-vel}. This is confined to weakly-interacting hit-and-runs resulting in two large remnants with debris that resides between them in space. These debris experience the gravitational potential of both remnants, which are in opposite directions. Thus, it is not a contradiction that the debris are not bound to either remnant while they have negative energy. These cases result in debris with low relative velocity (lower than that of the runner). 
This peculiar behavior should not be affecting the overall analysis, as they only occur in simulations with a low mass fraction of debris.
While the major characteristics of the largest remnants are well determined in this study, a more informed assessment of debris will require higher resolution simulations running to much later time. 

\section{Summary and Conclusion}
\label{sec:conclusion}

We present a new database of 1250 giant impact simulations relevant to terrestrial planet formation, spanning \SI{1e-6}{\mearth} to \SI{5}{\mearth} bodies, performed using SPH including friction. The dataset has six free parameters: the target mass, the impactor-to-target mass ratio, the core mass fractions of target and impactor, and the impact velocity and angle. Pre-impact rotation was not included. Some simulations were performed with a resolution increase by a factor of 10. Only small differences were found between our nominal resolution (100k SPH particles) and the high-resolution case (1M SPH particles).

In previous work, \citet{2019ApJCambioni}, \citet{2020ApJGabriel}, and \citet{2020ApJEmsenhuberA} reported on the accretion efficiencies, the change in composition, and the relative orbits of the largest remnants for giant impacts, based on a previous database \citep{2011PhDReufer}. As part of this new and more comprehensive database, we additionally calculate the moments of inertia of the final bodies, and determine the presence of major satellites. 

All simulations apply a solid friction model modulated by temperature-dependent yielding.
As expected, the effects of strength are most significant at smaller sizes. 
Friction acts to reduce differential flow velocities, and this can substantially change the gravitational and angular momentum dynamics.
Because of this, large satellites, several percents of the target mass, are found more commonly in the aftermath of collisions up to \SI{\sim2e-3}{\mearth}.
Earth-mass and larger collisions do not tend to feature captured impactors as satellites.
Low-mass bodies also have a diversity of extended, non-spherical post-impact shapes and interior structures.
These observed strength-supported features are consistent with previous studies on similar-sized collisions at smaller scales than those examined here \citep[e.g.,][]{2009PSSRichardson,2017AAJutziBenz}.
One possible example of such a strength supported structure is asteroid Pallas, over 500~km diameter, whose shape substantially deviates from equilibrium considering its current rotation \citep{marsset2020violent}.

We also present a systematic characterization of debris produced in giant impacts. We find that debris fragments are very small, in most cases, compared to the total mass of colliding bodies (less than \SI{1}{\percent}). Also, contrary to the common assumption \citep[e.g.,][]{2013IcarusChambers} that debris escape at low velocity, we find that the the mean debris velocity at infinity from giant impacts is about half of the relative velocity of the incoming objects, with a dispersion of $\sim$50\% from the mean. We did not find any systematic correlation between debris size and velocity, so an assumption that the two are independent is sufficient. Such simple scaling could help future \textit{N}-body-based evolution models to incorporate realistic debris dynamics.

\begin{acknowledgments}
We thank the two anonymous reviewers, whose comments and suggestions helped improve this manuscript.
A.E., E.A., S.R.S., and R.M. acknowledge support from NASA under grant 80NSSC19K0817 and the University of Arizona.
This work was funded by the Deutsche Forschungsgemeinschaft (DFG, German Research Foundation) - 362051796.
S.C. acknowledges support from the Crosby Distinguished Postdoctoral Fellowship of the Department of Earth, Atmospheric and Planetary Sciences of the Massachusetts Institute of Technology.
T.S.J.G. acknowledges support from the U.S. Geological Survey Astrogeology Science Center.
An allocation of computer time from the UA Research Computing High Performance Computing (HPC) is gratefully acknowledged.
Any use of trade, firm, or product names is for descriptive purposes only and does not imply endorsement by the U.S. Government. 
\end{acknowledgments}

\software{SPHLATCH \citep{2012IcarusReufer,2013IcarusAsphaug,2014NatGeoAsphaug,2018IcarusEmsenhuber,2023IcarusBallantyne}, matplotlib \citep{2007CSEHunter}, SciPy \citep{2020NatMethScipy}, NumPy \citep{2020NatureNumPy}}

\bibliographystyle{aasjournal}
\bibliography{manu,add}

\appendix

\section{1D calculations: bulk density and the moment of inertia}
\label{sec:1d}

\begin{figure*}
    \centering
    \includegraphics{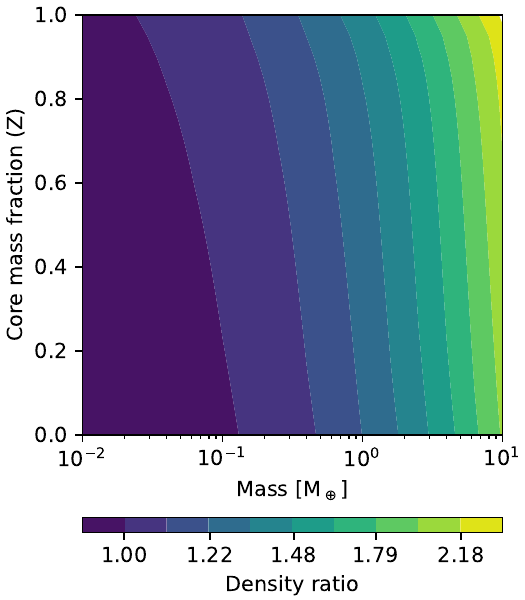}
    \includegraphics{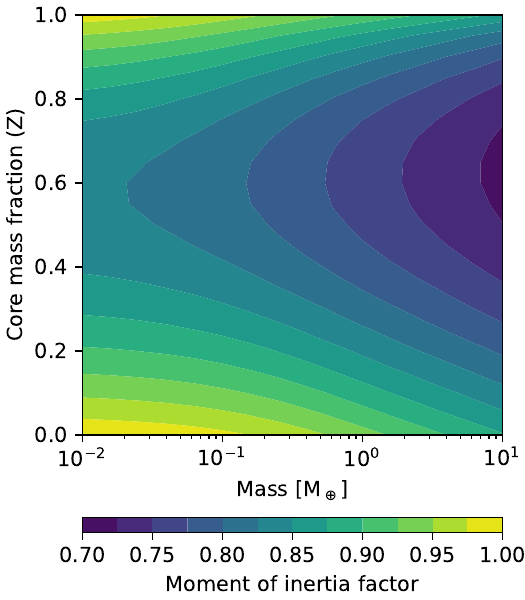}
    \caption{Density ratio with respect to the reference value from the EOS (left) and moment of inertia factor $I/I_\mathrm{ref}$ (such that 1 is a uniform density sphere; right) as function of body mass and core mass fraction.}
    \label{fig:denratio}
\end{figure*}

To ensure that the remnants can be treated consistently from pre-impact to post-impact in an \textit{N}-body simulation, we use the value for the radius that an initial body with the same mass and core mass fraction would have in our simulations. For this purpose, we compute a grid of 1D radial structures using the same approach as for the initialization of SPH collisions. Once these models are computed, we compute not only the bulk density (which is needed for the radius), but also other related properties, such as the moment of inertia. 

The results of these calculations are provided in Figure~\ref{fig:denratio}. To leave out the changes due to variable composition (the core mass fraction) and focus on the internal compression due to self-gravity, we show the ratio of the bulk density to the reference density from the EOS we are using for the SPH simulations: \texttt{ANEOS} for the iron core (with a reference density of \SI{7.65}{\gram\per\cubic\centi\meter}) and a modified version of the same \citep{2019AIPStewart} for forsterite in the mantle (with a reference density of \SI{3.32}{\gram\per\cubic\centi\meter}). For a given core-mass fraction $Z$, the reference density $\rho_\mathrm{ref}$ is computed as
\begin{equation}
    \left(\rho_\mathrm{ref}/\si{\gram\per\cubic\centi\meter}\right)^{-1}=(1/7.65)Z+(1/3.32)(1-Z).
\end{equation}
The calculations were performed for bodies down to \SI{e-5}{\mearth}. Figure~\ref{fig:denratio} shows only the results for bodies larger than \SI{e-2}{\mearth} because the smaller objects exhibit negligible compression. Their density is slightly below the reference value of the EOS because of the thermal expansion due to the internal energy. One outcome of the calculations is that the iron core is more compressible than the mantle in the EOSs that we use. This results in slightly different core-to-total radius ratios in bodies with different masses but otherwise the same core mass fraction.

The right panel of Figure~\ref{fig:denratio} depicts the ratio of the moment of inertia of the computed bodies with respect to the moment of inertia of an equivalent body with the same mass and radius, but assuming a constant density,
\begin{equation}
    I_\mathrm{ref} = \frac{2}{5}mr^2.
\end{equation}
Bodies with intermediate core mass fractions have the smallest values of the moment of inertia ratio, as expected. Nevertheless, the effect of the different compressibility of the core and mantle is also noticeable here, as the ratio is smaller for a given mass at $Z=1$ than at $Z=0$. We also computed the ratio of the potential energies, but since the results are very similar to the moment of inertia, they are not shown here. 

\section{Effect of the solid model parameters}
\label{sec:res-solidmodel}

\begin{figure*}
    \centering
    \makebox[3.5in]{\Large Nominal}
    \makebox[3.5in]{\Large Modified $\mu_\mathrm{d}$ and $\mu_\mathrm{i}$}
    \includegraphics[width=3.5in]{scatter_lhs_1000_058}
    \includegraphics[width=3.5in]{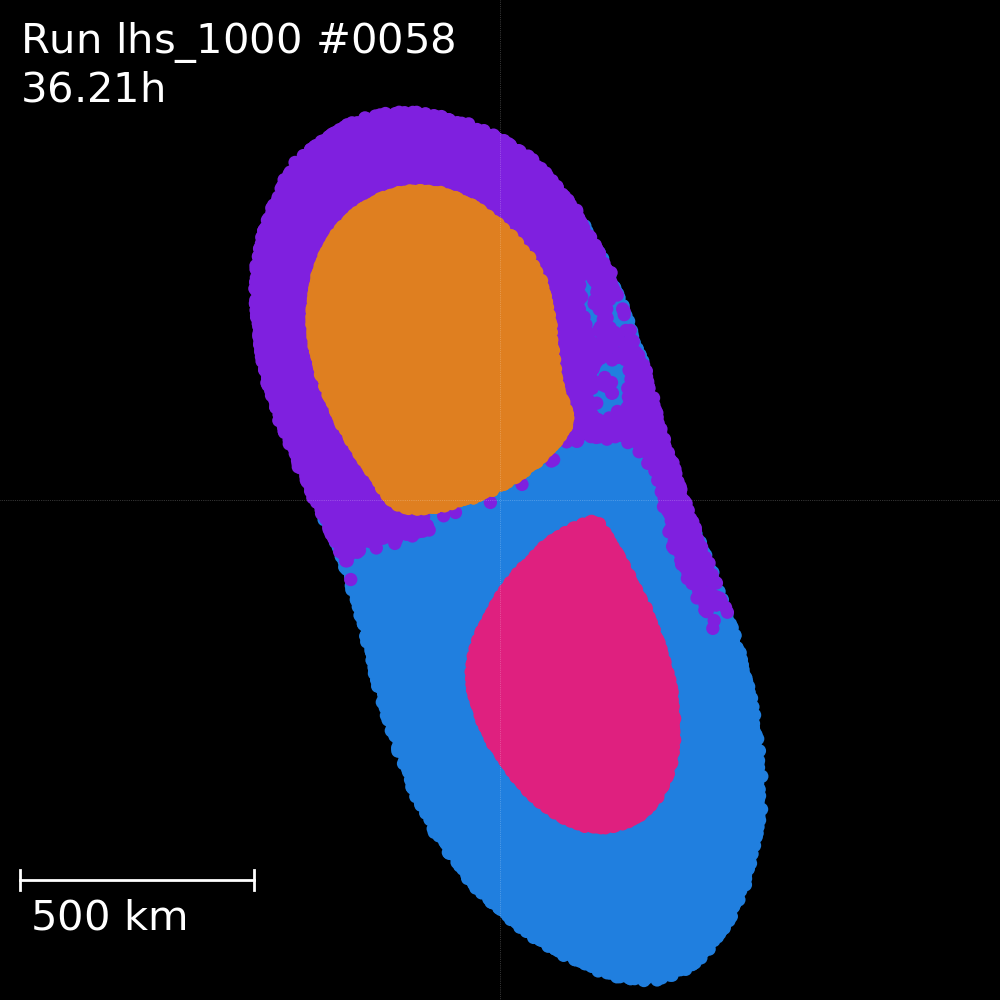}
    \caption{Final state of simulation \texttt{\#0058} with different values for parameters of the solid model. The left panel shows the nominal case and is the same as the right panel of Fig.~\ref{fig:scatter}. The right panel shows the same initial conditions, but with $\mu_\mathrm{d}=0.63$ and $\mu_\mathrm{i}=1.58$.}
    \label{fig:scater-lhs-1000-058}
\end{figure*}

To check how our choice for the values of $\mu_\mathrm{d}$ and $\mu_\mathrm{i}$ affect our results, we perform one simulation where we altered the values of $\mu_\mathrm{d}$ and $\mu_\mathrm{i}$. We selected simulation \texttt{\#0058} of \texttt{lhs\_1000}, as it one with the smallest bodies, lowest velocity, and oblique impact angle, which are those most affected by the inclusion of strength in our model. The nominal values for the parameters are $\mu_\mathrm{d}=0.8$ and $\mu_\mathrm{i}=2$, as discussed in Sect.~\ref{sec:strength}. For the altered values, we selected $\mu_\mathrm{d}=0.63$ and $\mu_\mathrm{i}=1.58$ following \citet{2016ScienceJohnson} and references therein.

The final state of the two simulations is provided in Figure~\ref{fig:scater-lhs-1000-058}. The two show essentially the same outcome with an elongated body and separate cores. Differences at smaller scales nevertheless appear, such as an even more elongated shape and thinner separation between the cores in the simulation with altered parameters. Thus, we find that the precise choice of the values for $\mu_\mathrm{d}$ and $\mu_\mathrm{i}$ parameters of the solid model does not affect the global outcome of the simulations. The results presented in this work are therefore robust in this regard.

\section{Effect of SPH resolution}
\label{sec:res-res}

The resolution of SPH simulations is set by the number of particles (resolution) in the simulation. To check whether our results are affected by resolution, we performed five additional runs with different number of particles but otherwise identical SPH parameters. For this analysis we predominantly focus on metrics related to debris, namely the clumps or fragments that are made of relatively few SPH particles.

\subsection{Global results}

\begin{table*}
\movetabledown=2.9in
\begin{rotatetable*}
    \centering
    \caption{Intermediate outcomes of five representative collision scenarios computed at two different SPH resolutions, $N=$\num{e5} (from the database) and $N=$\num{e6}, shown after first contact. Intermediate outcomes are especially useful for comparing the graze-and-merge cases where the bodies recollide later in the simulation. Each group of two rows represents one scenario simulated at both resolutions: graze-and-merge, hit-and-run, another graze-and-merge, near-head-on erosive, and slightly off-axis highly erosive/disruptive.  }
    \label{table:sphresf}
    \begin{tabular}{ccccccc|cccccccccccccc}
    \hline
    \multicolumn{7}{c|}{Collision Parameters} & \multicolumn{14}{c}{After 1\textsuperscript{st} enc.} \\
    $m_\mathrm{tar}$ & $\gamma$ & $Z_\mathrm{tar}$ & $Z_\mathrm{imp}$ & $\frac{v_\mathrm{coll}}{v_\mathrm{esc}}$ &  $\theta_\mathrm{coll}$ & $N$ & $\xi_\mathrm{L}$ & $\xi_\mathrm{S}$ & $\xi_\mathrm{D}$ & $\Omega_\mathrm{L}$ & $Z_\mathrm{L}$ & $I_\mathrm{Z,L}$ & $I_\mathrm{A,L}$ & $\epsilon'$ & $b'$ & $\Delta\varpi$ & $\Omega_\mathrm{S}$ & $Z_\mathrm{S}$ & $I_\mathrm{Z,S}$ &$I_\mathrm{A,S}$ \\
    $[\si{\mearth}]$ & & & & & [deg] & & & & & [rad/hr] & & & & & & [rad] & [rad/hr] & & \\ 
    \hline
    \hline
    \num{1.72e-04} & \num{0.961} & \num{0.371} & \num{0.284} & \num{1.169} & \num{51.2} & \num{e5} & \num{0.004} & \num{-0.004} & \num{0.000} & \num{0.840} & \num{0.369} & \num{1.127} & \num{1.043} & \num{-0.196} & \num{0.892} & \num{-0.687} & \num{0.834} & \num{0.281} & \num{1.135} & \num{1.047} \\
    \num{1.72e-04} & \num{0.961} & \num{0.371} & \num{0.284} & \num{1.169} & \num{51.2} & \num{e6} & \num{0.004} & \num{-0.004} & \num{0.000} & \num{0.734} & \num{0.368} & \num{1.114} & \num{1.036} & \num{-0.147} & \num{0.881} & \num{-0.640} & \num{0.717} & \num{0.281} & \num{1.111} & \num{1.037} \\
    \hline
    \num{7.14e-04} & \num{0.319} & \num{0.192} & \num{0.294} & \num{1.837} & \num{30.3} & \num{e5} & \num{0.209} & \num{-0.321} & \num{0.112} & \num{0.844} & \num{0.200} & \num{1.107} & \num{1.039} & \num{0.334} & \num{0.719} & \num{-0.979} & \num{1.579} & \num{0.312} & \num{1.174} & \num{1.063} \\
    \num{7.14e-04} & \num{0.319} & \num{0.192} & \num{0.294} & \num{1.837} & \num{30.3} & \num{e6} & \num{0.208} & \num{-0.354} & \num{0.146} & \num{0.764} & \num{0.201} & \num{1.108} & \num{1.032} & \num{0.350} & \num{0.699} & \num{-0.994} & \num{1.239} & \num{0.324} & \num{1.120} & \num{1.045} \\
    \hline
    \num{1.01e-03} & \num{0.052} & \num{0.398} & \num{0.130} & \num{1.026} & \num{51.7} & \num{e5} & \num{0.464} & \num{-0.469} & \num{0.005} & \num{0.318} & \num{0.393} & \num{1.024} & \num{1.016} & \num{-0.261} & \num{0.940} & \num{-1.030} & \num{6.720} & \num{0.160} & \num{11.162} & \num{7.585} \\
    \num{1.01e-03} & \num{0.052} & \num{0.398} & \num{0.130} & \num{1.026} & \num{51.7} & \num{e6} & \num{0.412} & \num{-0.414} & \num{0.002} & \num{0.273} & \num{0.394} & \num{1.013} & \num{1.009} & \num{-0.230} & \num{0.882} & \num{-1.035} & \num{10.048} & \num{0.158} & \num{20.187} & \num{13.636} \\
    \hline
    \num{4.80e-02} & \num{0.812} & \num{0.731} & \num{0.270} & \num{2.225} & \num{12.5} & \num{e5} & \num{-0.081} & \num{-0.764} & \num{0.844} & \num{0.713} & \num{0.725} & \num{1.049} & \num{1.214} & \num{-0.392} & \num{0.245} & \num{-2.556} & \num{1.023} & \num{0.623} & \num{7.755} & \num{5.539} \\
    \num{4.80e-02} & \num{0.812} & \num{0.731} & \num{0.270} & \num{2.225} & \num{12.5} & \num{e6} & \num{-0.145} & \num{-0.725} & \num{0.870} & \num{0.591} & \num{0.724} & \num{0.954} & \num{1.241} & \num{-0.423} & \num{0.262} & \num{-2.538} & \num{0.632} & \num{0.626} & \num{1.488} & \num{1.273} \\
    \hline
    \num{3.42e-01} & \num{0.458} & \num{0.542} & \num{0.520} & \num{5.210} & \num{23.5} & \num{e5} & \num{-1.249} & \num{-1.000} & \num{2.249} & \num{0.018} & \ldots & \num{1.853} & \num{1.733} & \ldots & \ldots & \ldots & \ldots & \ldots & \ldots & \ldots \\
    \num{3.42e-01} & \num{0.458} & \num{0.542} & \num{0.520} & \num{5.210} & \num{23.5} & \num{e6} & \num{-1.263} & \num{-1.000} & \num{2.263} & \num{0.040} & \ldots & \num{1.686} & \num{1.636} & \ldots & \ldots & \ldots & \ldots & \ldots & \ldots & \ldots \\
    \end{tabular}
\end{rotatetable*}
\end{table*}

\begin{table*}
\movetabledown=2.9in
\begin{rotatetable*}
    \centering
    \caption{The final outcomes of the same five collision scenarios in Table~\ref{table:sphresf}, again as a function of SPH resolution.}
    \label{table:sphrese}
    \begin{tabular}{ccccccc|ccccccccccc}
    \hline
    \multicolumn{7}{c|}{Collision parameters} & \multicolumn{11}{c}{End state} \\
    $m_\mathrm{tar}$ & $\gamma$ & $Z_\mathrm{tar}$ & $Z_\mathrm{imp}$ & $\frac{v_\mathrm{coll}}{v_\mathrm{esc}}$ &  $\theta_\mathrm{coll}$ & $N$ & $\xi_\mathrm{L}$ & $\xi_\mathrm{S}$ & $\xi_\mathrm{D}$ & $\Omega_\mathrm{L}$ & $Z_\mathrm{L}$ & $I_\mathrm{Z,L}$ & $I_\mathrm{A,L}$ & $\xi_\mathrm{P}$ & $\gamma_\mathrm{P}$ & $\frac{v_\mathrm{RMS}}{v_\mathrm{esc}}$ & $\frac{v_\mathrm{RMS,\infty}}{v_\mathrm{esc}}$ \\
    $[\si{\mearth}]$ & & & & & [deg] & & & & & [rad/hr] & & & & & & & \\ 
    \hline
    \hline
    \num{1.72e-04} & \num{0.961} & \num{0.371} & \num{0.284} & \num{1.169} & \num{51.2} & \num{e5} & \num{1.000} & \num{-1.000} & \num{0.000} & \num{4.366} & \num{0.326} & \num{2.474} & \num{1.801} & \num{1.000} & \num{0.000} & \ldots & \ldots \\
    \num{1.72e-04} & \num{0.961} & \num{0.371} & \num{0.284} & \num{1.169} & \num{51.2} & \num{e6} & \num{1.000} & \num{-1.000} & \num{0.000} & \num{4.367} & \num{0.326} & \num{2.394} & \num{1.746} & \num{1.000} & \num{0.000} & \ldots & \ldots \\
    \hline
    \num{7.14e-04} & \num{0.319} & \num{0.192} & \num{0.294} & \num{1.837} & \num{30.3} & \num{e5} & \num{0.209} & \num{-0.321} & \num{0.112} & \num{0.844} & \num{0.200} & \num{1.107} & \num{1.039} & \num{0.209} & \num{0.203} & \num{0.838} & \num{0.828} \\
    \num{7.14e-04} & \num{0.319} & \num{0.192} & \num{0.294} & \num{1.837} & \num{30.3} & \num{e6} & \num{0.208} & \num{-0.354} & \num{0.146} & \num{0.764} & \num{0.201} & \num{1.108} & \num{1.032} & \num{0.165} & \num{0.179} & \num{0.956} & \num{0.947} \\
    \hline
    \num{1.01e-03} & \num{0.052} & \num{0.398} & \num{0.130} & \num{1.026} & \num{51.7} & \num{e5} & \num{0.917} & \num{-0.928} & \num{0.011} & \num{0.534} & \num{0.389} & \num{1.040} & \num{1.018} & \num{0.875} & \num{0.004} & \num{0.561} & \num{0.543} \\
    \num{1.01e-03} & \num{0.052} & \num{0.398} & \num{0.130} & \num{1.026} & \num{51.7} & \num{e6} & \num{0.894} & \num{-0.942} & \num{0.048} & \num{0.506} & \num{0.388} & \num{1.032} & \num{1.011} & \num{0.836} & \num{0.003} & \num{0.357} & \num{0.320} \\
    \hline
    \num{4.80e-02} & \num{0.812} & \num{0.731} & \num{0.270} & \num{2.225} & \num{12.5} & \num{e5} & \num{0.278} & \num{-0.953} & \num{0.675} & \num{1.559} & \num{0.696} & \num{1.410} & \num{1.268} & \num{0.206} & \num{0.032} & \num{1.050} & \num{1.078} \\
    \num{4.80e-02} & \num{0.812} & \num{0.731} & \num{0.270} & \num{2.225} & \num{12.5} & \num{e6} & \num{0.248} & \num{-0.923} & \num{0.675} & \num{1.570} & \num{0.697} & \num{1.236} & \num{1.176} & \num{0.170} & \num{0.052} & \num{1.134} & \num{1.127} \\
    \hline
    \num{3.42e-01} & \num{0.458} & \num{0.542} & \num{0.520} & \num{5.210} & \num{23.5} & \num{e5} & \num{-1.249} & \num{-1.000} & \num{2.249} & \num{0.018} & \num{0.868} & \num{1.853} & \num{1.733} & \num{-1.249} & \num{0.000} & \num{2.398} & \num{2.395} \\
    \num{3.42e-01} & \num{0.458} & \num{0.542} & \num{0.520} & \num{5.210} & \num{23.5} & \num{e6} & \num{-1.263} & \num{-1.000} & \num{2.263} & \num{0.040} & \num{0.868} & \num{1.686} & \num{1.636} & \num{-1.263} & \num{0.000} & \num{2.403} & \num{2.400} \\
    \end{tabular}
\end{rotatetable*}
\end{table*}

We provide a comparison of the global outcomes five simulations in Table~\ref{table:sphrese} and Table~\ref{table:sphresf}. The former represents outcomes after the first contact (\emph{i.e.}, the first phase of a graze-and-merge collisions) and the latter represents the final outcome. The columns provided are the same as those in Table~\ref{table:sph}, except they include a new parameter, $N$, for the SPH resolution (particle count), and we removed the outcome $n_\mathrm{P}$ since satellites are not produced in these collisions. Furthermore, we did not include the number of resolved clumps $N_\mathrm{res}$ in these tables, as we analyze this later in this section.

The simulations for this resolution study were selected to represent a variety of outcomes from the overall database. Most of the metrics that we are studying are similar whether the number of particles $N$ is \num{e5} or \num{e6}. The perfect merger collision, shown in the first two rows in each table, shows nearly identical results in all outcomes as a function of resolution. For the others, accretion efficiencies show differences in the range of \num{0.01} to \num{0.03}, which can equate roughly to less than a few percent of total mass depending on the impactor mass. The final rotation rate and core mass fraction of the largest remnant are also very similar across resolution in the final state. For the intermediate state, rotation rate of the largest remnant seems to be somewhat variable, showing $\sim$20\% relative deviation in some cases. Debris velocities are also more uncertain, in two cases with differences of up to $0.2 v_\mathrm{esc}$. These differences, however, do not generally exceed the statistical variations of these quantities in the database under small variations in impact angle or velocity. Additionally, as noted, in cases where the core mass fractions of the initial bodies are very large, the mantle materials are less well resolved, and thus may be subject to SPH challenges resolving material discontinuity across the core-mantle boundary. While these cases might be expected to show greater variations as a function of $N$, in our resolution comparisons with the largest core mass fractions we do not observe large deviations in the composition or other properties of the remnants. Thus, while higher resolution studies produce some differences, that are important to explore further as resources permit, overall we conclude that our findings are robust at the $N=\num{e5}$ resolution of the database.

\subsection{Size-frequency distribution}
\label{sec:res-ressfd}

\begin{figure}
    \centering
    \includegraphics{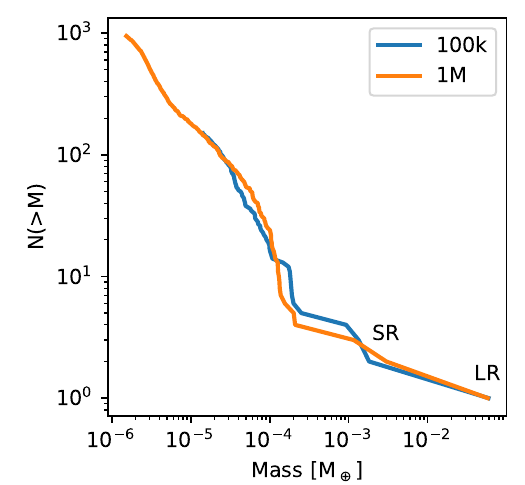}
    \caption{Comparison of the size-frequency distribution of the remnants for collision \texttt{566} of the 1000-simulation set for two different SPH resolutions, as indicated in the legends. The two texts ``LR'' and ``SR'' denote the largest and second remnants, respectively.}
    \label{fig:debris-sfd-res}
\end{figure}

\begin{figure*}
    \centering
    \includegraphics{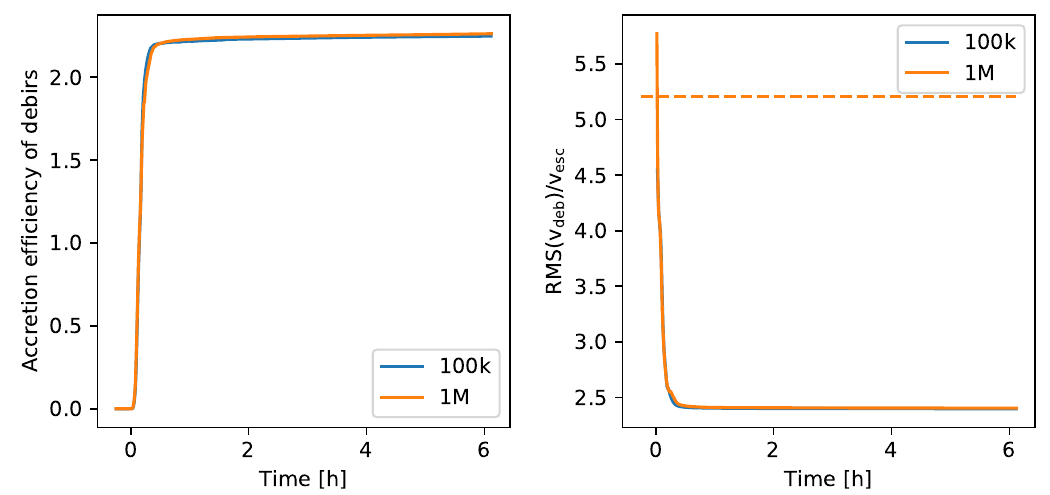}
    \caption{Comparison of the accretion efficiency of debris $\xi_\mathrm{D}$ and the RMS of the debris velocity for collision \texttt{561} of the 1000-simulation set for two different SPH resolutions, as indicated in the legends. The dashed horizontal line in the right panel shows the impact velocity.}
    \label{fig:orb-lhs-1000-0561}
\end{figure*}

To check for the robustness of the size-frequency distribution with resolution, we selected one simulation where several hundreds of clumps are produced as debris: simulation \texttt{566}, with $m_\mathrm{tar}=\SI{4.80e-2}{\mearth}$, $\gamma=0.81$, $v_\mathrm{coll}/v_\mathrm{esc}=2.23$, and $\theta_\mathrm{coll}=\SI{12.5}{\degree}$. The size-frequency distribution at the end of the evolution for two simulations, one from the database (labeled ``100k'') and a similar one but with 1 million SPH particles instead (labeled ``1M'') are shown in Figure~\ref{fig:debris-sfd-res}. Overall, the two distributions appear to match quite well; the mass of the largest remnant is similar in both simulations and the total number of clumps at the resolution limit of the nominal simulation is also similar. There are nevertheless a few differences at intermediate remnant masses. The high-resolution simulation produces a more massive second remnant, as expected \citep[e.g.,][]{2015IcarusGenda}. The high-resolution run also only has two remnants comparable in mass to the second-largest, whereas the nominal-resolution simulation has three.

\subsection{Debris velocity}

To understand whether debris masses and velocities are influenced by common choices in SPH resolution used in giant impact studies, we perform a simulation at 10$^6$ particles and compare it to the nominal resolution in the dataset ($10^5$ particles). We select a run from the database where debris represents roughly \SI{70}{\percent} of the total mass, and where no clumps are found (simulation \texttt{561}, $m_\mathrm{tar}=\SI{3.42e-1}{\mearth}$, $\gamma=0.46$, $v_\mathrm{coll}/v_\mathrm{esc}=5.21$, $\theta_\mathrm{coll}=\SI{23.5}{\degree}$). As shown in Figure~\ref{fig:orb-lhs-1000-0561}, the amount of debris produced and their velocities are very similar in both simulations; the debris are traveling at roughly 2.5 times the mutual escape velocity of the colliding bodies, or nearly half of their relative velocity at initial contact. In addition, no clumps (minimum of 10 particles) are found within the debris even when using 1 million particles, so debris size remains unresolved.

\end{document}